\documentclass[12pt]{iopart}
\usepackage{cite}

\usepackage[usenames]{color}\definecolor{DkRed}{cmyk}{0,.5,.5,.3}

\def\Nfootnote#1#2{\footnote{#2}}

\date{Resubmitted September 2008}

\def\pnlabel#1{\label{#1}}
\def\includefigs{y}
\newcommand{\cut}[1]{\ERROR}
\def\arxivfigs{y}

\usepackage{times}
\usepackage{graphicx}
\usepackage{verbatim} 
\usepackage{bm}

\begin{document}
\title{First-principles calculation of DNA looping in tethered particle experiments}
\author{Kevin B Towles$^{1}$, John F Beausang$^{1}$, Hernan G Garcia$^{2}$, Rob Phillips$^{3}$, and Philip C Nelson$^{1}$}
\address{$^{1}$Department of Physics and Astronomy, University of  Pennsylvania, Philadelphia PA 19104}
\address{$^{2}$Department of Physics, California Institute of Technology, Pasadena CA 91125}
\address{$^{3}$Division of Engineering and Applied Science, California Institute of Technology, Pasadena CA 91125}
\ead{nelson@physics.upenn.edu}

\def\truexivalue{45} 
\def\wstarvalue{187\,\nmunit^2\msunit^{-1}}
\def\shortJvalues{$\approx80$, $350$, and $ 190\,\nMunit$}

\def\runit{\ensuremath{\mathsf{rad}}}
\def\gunit{\ensuremath{\mathrm g}}\def\mgunit{\ensuremath{\mathrm{mg}}}
\def\mlunit{\ensuremath{\mathrm{ml}}}
\def\mMunit{\ensuremath{\mbox{m\textsc m}}}
\def\nMunit{\ensuremath{\mbox{n\textsc m}}}\def\Munit{\ensuremath{\mbox{\textsc m}}}
\def\pMunit{\ensuremath{\mbox{p\textsc m}}}
\def\that{\hat t}
\def\msunit{\ensuremath{\mathrm{ms}}}
\def\pNunit{\ensuremath{\mathrm{pN}}}
\def\Junit{\ensuremath{\mathrm{J}}}
\def\bpunit{\ensuremath{\mathrm{bp}}}
\def\nmunit{\ensuremath{\mathrm{nm}}}
\def\umunit{\ensuremath{\mu\mathrm{m}}}
\def\sunit{\ensuremath{\mathrm{s}}}

\newcommand{\pitemize}{\begin{itemize}\setlength{\itemsep}{0pt}\setlength{\parsep}{0pt}}
\newcommand{\xpitemize}{\end{itemize}}
\newcommand{\penumerate}{\begin{enumerate}\setlength{\itemsep}{0pt}\setlength{\parsep}{0pt}}
\newcommand{\xpenumerate}{\end{enumerate}}
\def\Nitem#1{\par\hangindent=1.5em\noindent\small\hbox to 1.5em{\ {\bf#1$\bullet$ } }}
\def\Nlist#1{\par\hangindent=1.5em\noindent\small\hbox to 2em{\ {#1~} }}
\def\Nxitem{\par\normalsize\hangindent=0em\noindent}
\def\Nitemitem#1{\par\hangindent=2.5em\footnotesize\noindent\hbox to 2em{\ #1. }}
\def\Nxitemitem{\par\normalsize\smallskip\hangindent=1.5em}

\newcommand{\vr}{\mathbf{r}}
\newcommand{\eff}{_{\rm eff}}
\newcommand{\sloop}{_{{\rm loop}}}
\newcommand{\sun}{_{{\rm unl}}}
\newcommand{\sbead}{_{{\rm bead}}}
\newcommand{\stot}{_{\rm tot}}
\def\smax{_{{\rm max}}}
\def\kbt{k_{\rm B}T}
\newcommand{\Nlaci}{\ensuremath{\mathrm{[LacI]}}}
\def\Nrho{\rho}
\newcommand{\dna}{{\sffamily 3DNA }}\newcommand{\dnax}{{\sffamily 3DNA}}

\newcommand{\Nastate}{\ensuremath{\mbox{\bfseries\slshape a}}}\newcommand{\Nbstate}{\ensuremath{\mbox{\bfseries\slshape b}}}
\newcommand{\Nastatesmall}{\mbox{\small\bfseries\slshape a}}\newcommand{\Nbstatesmall}{\mbox{\small\bfseries\slshape b}}
\def\threethree#1#2#3{{\textstyle\left[{{#1\atop{\vphantom{q}#2}}\atop{\vphantom{I}\scriptscriptstyle#3}} \right]}}
\def\half{{\textstyle\frac12}}
\newcommand{\capitem}[1]{(\textit{\lowercase{#1}})~}
\def\exv#1{\langle #1\rangle}
\def\dd{\mathrm{d}}
\def\ex#1{{\rm e}^{#1}}
\newcommand{\ee}[1]{\cdot10^{#1}}
\newcommand{\eem}[1]{\cdot10^{-#1}}
\newcommand{\pnvec}{\mathbf}
\newcommand{\inv}{^{\raise.15ex\hbox{${\scriptscriptstyle-}$}\kern-.05em 1}}

\gdef\Npers{\xi}
\gdef\NRbead{R_{\rm bead}}\gdef\NLl{L_{\rm loop}}
\gdef\Nshutt{\delta t}
\gdef\Nviscos{\eta}
\gdef\Nsamp{N_{\rm samp}}
\gdef\Ntaudrift{\tau_{\rm diff}}
\gdef\Nen{{\cal E}}
\gdef\Qmat{\mathsf{Q}}\gdef\Tmat{\mathsf{T}}\gdef\Dmat{\mathsf{D}}\gdef\Jmat{\mathsf J}\gdef\hmat{\mathsf h}\gdef\Rmat{\mathsf R}
\gdef\Npconst{\gamma}
\gdef\Nbx{\hat{\mathbf E}_1}\gdef\Nby{\hat{\mathbf E}_2}\gdef\Nbz{\hat{\mathbf E}_3}\gdef\Nbgeneric#1{\hat{\mathbf E}_{#1}}
\gdef\Nmvec{\mathbf{m}}
\gdef\Nrrms{\rho_{\scriptscriptstyle\rm RMS}}\gdef\Nrrmsfour{\rho_{{\scriptscriptstyle\rm RMS}\rm,\,4\,\sunit}}\gdef\Nrrmst{\rho_{{\scriptscriptstyle\rm RMS},t}}
\gdef\NPlooptot{P(\mathrm{looped})}
\gdef\NKd{K_{\rm d}}\gdef\NK{\error}
\gdef\NOid{\ensuremath{\mathrm{O}_{\mathrm{id}}}}\gdef\NOone{\ensuremath{\mathrm{O}_{1}}}
\gdef\Nlaci{\ensuremath{[\mathrm{LacI}]}}
\gdef\Ndv{\delta v}\gdef\Ndo{\delta\omega}
\gdef\NJ{J}\gdef\NJt{\widetilde J}\def\NJb{\bar J}
\gdef\NM{\mbox{\textsf{\bfseries M}}}\gdef\NN{\mbox{\textsf{\bfseries N}}}
\gdef\NNgen{N_{\rm gen}}
\gdef\NDG{\Delta G_{\rm loop}}
\def\tot{_{\mathrm{tot}}}
\def\Nspacing{I}
\def\Nbeadcount{N_{\rm beads}}

\def\eref#1{eqn.\thinspace(\ref{e:#1})}\def\erefs#1{eqns.\thinspace(\ref{e:#1}}
\def\erefsn#1{eqns.\thinspace\ref{e:#1}} 
\def\Eref#1{Equation\thinspace(\ref{e:#1})}
\def\sref#1{section\thinspace\ref{s:#1}}\def\Sref#1{Section\thinspace\ref{s:#1}}
\def\srefs#1{sections\thinspace\ref{s:#1}}
\def\fref#1{figure\thinspace\ref{f:#1}}\def\frefn#1#2{figure\thinspace\ref{f:#1}\textit{\lowercase{#2}}}
\def\Fref#1{Figure\thinspace\ref{f:#1}}\def\Frefn#1#2{Figure\thinspace\ref{f:#1}\textit{\lowercase{#2}}}
\def\tref#1{Table\thinspace\ref{t:#1}}
\def\frefs#1{figures\thinspace\ref{f:#1}}\def\Frefs#1{Figures\thinspace\ref{f:#1}}
\def\Nrref#1{ref.{}\thinspace\relax[\citenum{#1}]} 
\def\NRref#1{Ref.{}\thinspace\relax[\citenum{#1}]} 
\def\Nrrefs#1{refs.{}\thinspace\relax[\citenum{#1}]}
\def\NRrefs#1{Refs.{}\thinspace\relax[\citenum{#1}]}
\def\aref#1{section\thinspace{}\ref{s:#1}}\def\arefs#1{sections\thinspace{}\ref{s:#1}}
\def\Aref#1{Section\thinspace{}\ref{s:#1}}\def\Arefs#1{Sections\thinspace{}\ref{s:#1}}

\def\yesflag{y}
\def\ifig#1#2#3{\begin{figure}[tb!]
\begin{minipage}[b]{.95\textwidth}
\ifx\includefigs\yesflag
\begin{center}
\ifx\arxivfigs\yesflag
\includegraphics[width=4truein]
{JPGtoEPSgraphics/#3}
\else\includegraphics
{Graphics/#3}\fi\end{center}\fi
 \end{minipage}
\hfil\par\smallskip \caption{\small #2 \pnlabel{f:#1}}
\end{figure}%
}
\def\ifigsize#1#2#3#4{\begin{figure}[tb!]
\begin{minipage}[b]{.95\textwidth}
\ifx\includefigs\yesflag
\begin{center}
\ifx\arxivfigs\yesflag
\includegraphics[width=#4truein]{JPGtoEPSgraphics/#3}
\else\includegraphics[width=#4truein]{Graphics/#3}\fi\end{center}\fi
 \end{minipage}
\hfil\par\smallskip \caption{\small #2 \pnlabel{f:#1}}
\end{figure}%
}

\begin{abstract}
We calculate the probability of DNA loop formation mediated by regulatory proteins such as {Lac} repressor (LacI), using a mathematical model of DNA elasticity. Our model is adapted to calculating
quantities directly observable in Tethered Particle Motion (TPM) experiments, and it accounts for all the entropic forces present in such experiments. Our model has {\it no free parameters}; it
characterizes DNA elasticity using information obtained in other kinds of experiments. It assumes a harmonic elastic energy function (or wormlike chain type elasticity), but our Monte Carlo
calculation scheme is flexible enough to accommodate arbitrary elastic energy functions. We show how to compute both the ``looping $J$ factor'' (or equivalently, the looping free energy) for various
DNA construct geometries and LacI concentrations, as well as the detailed probability density function of bead excursions. We also show how to extract the same quantities from recent experimental data
on tethered particle motion, and then compare to our model's predictions. In particular, we present a new method to correct observed data for finite camera shutter time and other experimental effects.

Although the currently available experimental data give large uncertainties, our first-principles predictions for the looping free energy change are confirmed to within about $1\,\kbt$, for loops of
length around $300\,$basepairs. More significantly, our model successfully reproduces the detailed distributions of bead excursion, including their surprising three-peak structure, without any fit
parameters and without invoking any alternative conformation of the LacI tetramer. Indeed, the model qualitatively reproduces the observed dependence of these distributions on tether length (e.g.,
phasing) and on LacI concentration (titration). However, for short DNA loops (around $95\,$basepairs) the experiments show more looping than is predicted by the harmonic-elasticity model, echoing
other recent experimental results. Because the experiments we study are done \textit{in vitro,} this anomalously high looping cannot be rationalized as resulting from the presence of DNA-bending
proteins or other cellular machinery. We also show that it is unlikely to be the result of a hypothetical ``open'' conformation of the LacI tetramer.
\end{abstract}

\submitto{\PB}

\section{Introduction and summary\pnlabel{s:is}}
\subsection{Background\pnlabel{s:b}}
Living cells must orchestrate a multitude of biochemical processes. Bacteria, for example, must rigorously suppress any unnecessary activities to maximize their growth rate, while maintaining the
potential to carry out those activities should conditions change. For example, in a glucose-rich medium {\it E.\ coli} turn off the deployment of the machinery needed to metabolize lactose; when
starved of glucose, but supplied with lactose, they switch this machinery on. This switch mechanism---the ``\textit{lac} operon''---was historically the first genetic regulatory system to be discovered. Physically, the
mechanism involves the binding of a regulatory protein,  called LacI, to a specific sequence of DNA (the ``operator'') situated near the beginning of the set of genes coding for the lactose
metabolism enzymes. Some recent reviews of the \textit{lac} system include \Nrrefs{Matthews1992,mull96a,matt98a,lewi05a}; see also \Nrref{Ptashne2004} for looping in the \textit{lambda} system.

Long after the discovery of genetic switching, it was found that some regulatory proteins, including LacI, exist in multimeric forms with two binding heads for DNA, and that their normal operation
involves binding both sites to distant operators, forming a loop \cite{Dunn1984,moss86a,kram87a,Hsieh1987,Adhya1989,reve99a}. The looping mechanism seems to confer advantages in terms of function
\cite{vila03a}. From the biophysical perspective, it is remarkable that in some cases
loop formation, and its associated gene repression, proceed \textit{in vivo} even when the distance between operators is much less than a persistence length of DNA \cite{Muller1996}. For this and
other reasons, a number of experimental methods have been brought to bear on reproducing DNA looping \textit{in vitro}, to minimize the effects of unknown factors and focus on the one process of
interest.  Reconstituting DNA looping behavior in this way is an important step in clarifying the mechanism of gene regulation.

Tethered particle motion (TPM) is an attractive technique for this purpose \cite{finz95a}. In this method, a long DNA construct is prepared with two (or more) operator sequences at a desired spacing near the
middle. One end is anchored to a wall, and the other to an otherwise free, optically visible bead. The bead motion is passively monitored, typically by tracking microscopy, and used as an indirect reporter of
conformational changes in the DNA, including loop formation and breakdown (\fref{cartoon}).
\ifigsize{cartoon}{\capitem{a}Cartoon of a DNA molecule flexibly linking a bead to a surface via freely pivoting attachments (not to scale). The motion of the bead's center is observed and tracked,
for example as described in \Nrref{nels06a}. In each video frame, the position vector, usually projected to the $xy$ plane, is found. After drift subtraction, the mean of this position vector defines
the anchoring point. The projected distance from this anchoring point to the instantaneous bead center is the bead excursion $\Nrho$.
A regulatory protein, for example a LacI tetramer, is shown bound to a specific ``operator'' site on the DNA.
\capitem{b}The conformational change of interest to us is loop formation: A loop forms when the repressor also binds to a second operator. The figure shows an actual representative
looped configuration from the simulations described in this paper, drawn to scale. \Frefs{FrameNotation} and \ref{f:goodbad} explain the graphical representations of DNA and LacI used here.
}{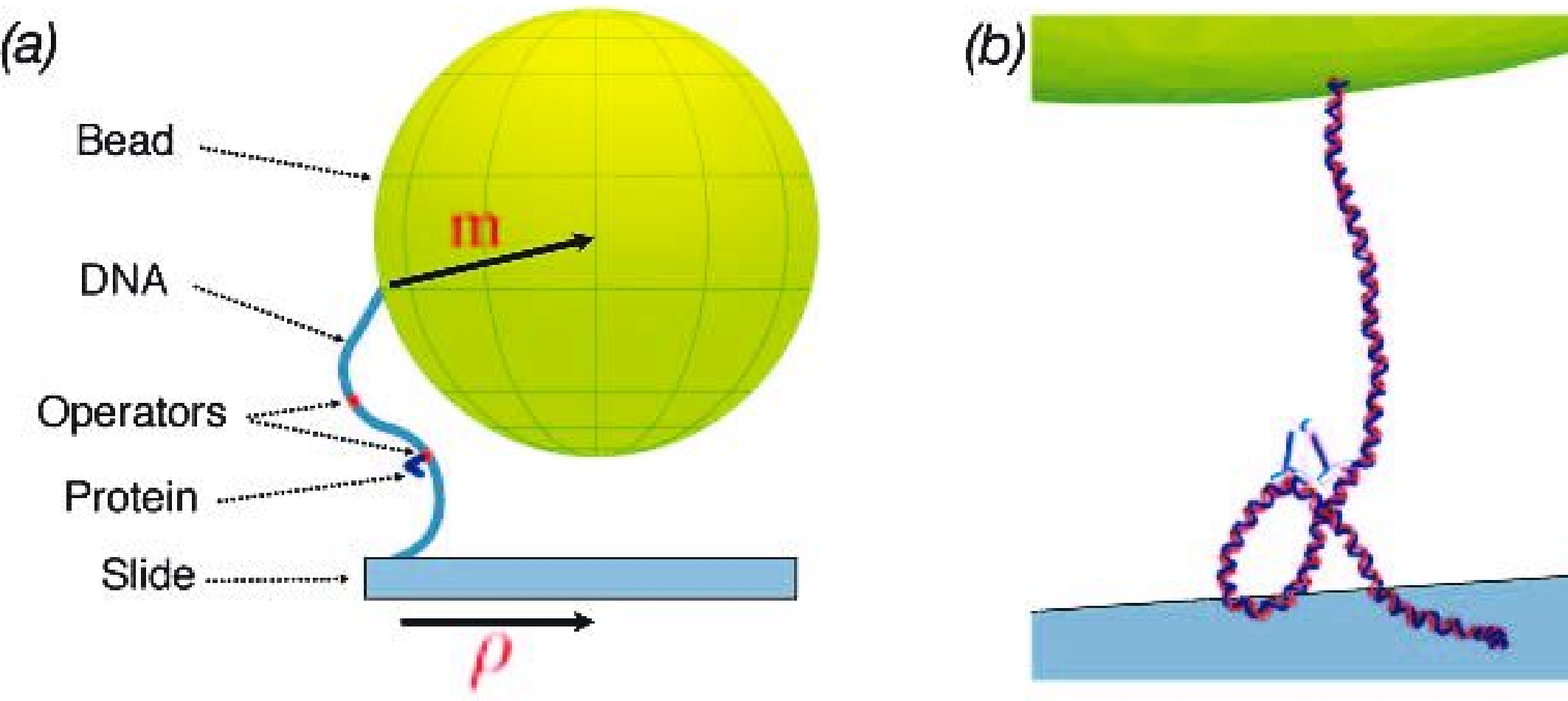}{4}

\subsection{Goals of this paper\pnlabel{s:gtp}}
The recent surge of interest in DNA looping motivated us to ask: Can we understand TPM data quantitatively, starting from simple models of DNA elasticity? What is the simplest model
that captures the main trends? How well can we predict data from TPM experiments, using {\it no fitting parameters?}

To answer such questions, we had to combine and improve a number of existing calculation tools. This paper explains how to obtain a simple elastic-rod model for DNA, and a geometric characterization
of the repressor--DNA complex, from existing (non-TPM) experiments. From this starting point, with no additional fitting parameters, we show how to calculate experimentally observable quantities of
TPM experiments (such as the fraction of time spent in various looped states and the distribution of bead excursions), as functions of experimentally controlled parameters (operator separation and
repressor concentration), and compare to recent experiments.

Although our main interest is TPM experiments, our method is more generally applicable. Thus as a secondary project, we also compute looping $J$ factors for a DNA construct with no bead or wall
(``pure looping''). This situation is closer to the one that prevails \textit{in vivo}; although in that case many other uncertainties enter, it is nevertheless interesting to compare our results to
the experimental data.

\subsection{Assumptions, methods, and results of this paper\pnlabel{s:rtp}}
Supplementary Information \aref{sn} gives a summary of  the notation used in this paper. Some readers may wish to skip to \sref{ior}, where we summarize our results. \Sref{oop} gives an outline of
the main text and the supplement; in addition, the other subsections of this introduction give forward references showing where certain key material can be found.

\subsubsection{Outline of assumptions}
First we summarize key assumptions and simplifications made in our analysis. Some will be justified in the main text, whereas others are taken in the spirit of seeking the model that is ``as simple as possible,
but not more so.''

All our results are obtained using equilibrium statistical mechanics; we make no attempt to obtain rate constants, although these are experimentally available from TPM data \cite{finz95a,vanz06a,beau07a,beau07b}.
Our model treats DNA as a homogeneous, helical, elastic body, described by a $3\times3$ elastic compliance matrix (discussed in \sref{etu}). Thus we neglect, for now, the effect of DNA sequence information, so our results may
be compared only to experiments done with random-sequence DNA constructs. Despite this reduction, our model is more realistic than ones
that have previously been used for TPM theory; for example, we include the substantial bend anisotropy, and twist--bend coupling, of DNA elasticity.
We also neglect long-range electrostatic interactions (as is appropriate at the high salt conditions in the experiments we study),
assuming that electrostatic effects can be summarized in effective values of the elastic compliances.

The presence of a large reporter bead at one end of the DNA construct, and a wall at the other end, significantly perturb looping in TPM experiments. We treat the bead as a sphere, the wall as a
plane, and the steric exclusion between them as a hard-wall interaction.

We treat DNA--protein binding at each of the two binding sites on the repressor as independent second-order reactions; that is, we assume no allosteric cooperativity. Moreover, we neglect nonspecific
DNA--protein interactions (``wrapping'' \cite{tsod99a}).

\subsubsection{Outline of methods}
Our method builds on prior work \cite{sega06a,nels06a}. \Sref{rtor} discusses other theoretical approaches in the literature.

Our calculations must include the effects of chain entropy on loop formation, because we consider loop lengths as large as 510~basepairs. We
must also account for entropic-force effects created by the large bead at one end of the DNA and the wall at the other end, in addition to the specific orientation constraints imposed on the two
operators by the repressor protein complex. To our knowledge such a complete, first-principles approach to calculating DNA looping for tethered particle motion has not previously been attempted. In part
because of these complications, we chose to calculate using a Monte Carlo method called ``Gaussian sampling'' (discussed in \sref{ctdwl} and \arefs{acg}--\ref{s:MCa}). Gaussian sampling is distinguished from Markov-chain methods (e.g.,\
Metropolis Monte Carlo) in that successive sampled chains are independent of their predecessors.

We must also address a number of points before we can compare our results to experiments. For example, DNA simulations
report a quantity called the ``looping $J$ factor.'' But TPM experiments instead report the time spent in looped versus unlooped states, which depends on both $J$ and a binding constant
$\NKd$. We present a method to extract both $J$ and $\NKd$ separately from TPM data (discussed in \aref{edlJf}). We also describe two new data-analysis tools: {\it(1)}\/~A
correction to our theoretical results on bead excursion, needed to
account for the effect of finite camera shutter time on the experimental results (discussed in \aref{bc2}), and {\it(2)}\/~another correction needed to make contact with a widely used statistic, the finite-sample RMS
bead excursion (discussed in \aref{sse}).
({To be precise, the latter two corrections do both involve phenomenological parameters, but we obtain these from TPM data that are different from the ones we are
seeking to explain. Each correction could in any case be avoided by taking the experimental data differently, as described in the Supplementary Information.})

\subsubsection{Outline of results\pnlabel{s:ior}}
Some of our results were first outlined in \Nrrefs{nels07a,nels08x}. The assumptions sketched above amount to a highly reductionist approach to looping. Moreover, we have given ourselves no freedom to tweak
the model with adjustable parameters, other than the few obtained from non-TPM experiments (four elastic constants and the geometry of the repressor tetramer); all other parameters we used had known
values (e.g., bead size and details of the DNA construct). So it is not surprising that some of our results are only in qualitative agreement with experiment. Nevertheless, we  find that:
\Nitem{}Our physical model quantitatively predicts basic aspects of the TPM experiments, such as the effects of varying tether length and bead size (see \fref{calibrplot}).
\Nitem{}The model can roughly explain the overall value of the looping $J$ factor obtained in experiments for a range of loop lengths near 300~basepairs (discussed in \sref{dl}).
\Nitem{}Perhaps most surprising, the same simple model predicts rather well the observed,
detailed structure of the \textit{distribution} of bead excursions, including its dependence on loop lengths near 300~basepairs
(see  e.g.\ \fref{FitLong}). The distinctive three-peaks structure of this distribution \cite{wong07a,norm08a,han?cLength} has sometimes been taken as \textit{prima facie} evidence for
a hypothetical alternate ``open'' conformation of the repressor protein. But we show that it can also arise without that hypothesis, as a consequence of the contributions of loops with different topologies.
\Nitem{}Notwithstanding those successes, our simple model does  \textit{not} successfully extrapolate to predict the magnitude of the $J$ factor for loop lengths near 100~basepairs, at least
according to the limited, preliminary experimental data now available.
Instead, there it  underestimates  $J$,  pointing to a breakdown of some of its hypotheses in this high-strain situation. Perhaps the needed modification is
a nonlinear elastic theory of DNA \cite{yan04a,wigg05a}, significant flexibility in the tetramer, additional nonspecific binding of DNA to the repressor protein,
or some combination of these.
\Nitem{}However, our model does give a reasonable account of the structure of the bead excursion distribution even for loop lengths near 100$\,\bpunit$ (see \fref{FitShort}).
\Nitem{}Because previous authors have proposed the specific hypothesis that one of the excursion-distribution peaks reflects an ``open'' conformation of LacI, we simulated that situation
as well. We argue that this hypothesis cannot by itself explain the high degree of looping observed experimentally for short DNA constructs (discussed in \sref{olrc}).
\Nxitem
Our calculations also quantify the importance of the orientation constraint for binding to the tetramer, via a concept we call the ``differential $J$ factor'' (discussed in \sref{lJf1}). Finally, our simple model
of blur correction quantitatively predicts the observed dependence of apparent bead motion on camera shutter time, and we expect it will be useful for future TPM experiments (discussed in \aref{bc}).

\subsubsection{Organization of this paper\pnlabel{s:oop}}
\Sref{stel} gives an overview of various single-molecule experiments used recently to study looping, emphasizing the particular capabilities of TPM.
\Sref{etu} derives the elastic model of DNA to be used in this paper.
\Sref{ctdwl} introduces our Monte Carlo method, and gives a crucial check that theory and experiment are both working properly,
by showing to what extent we can accurately predict the excursion of the tethered bead in the absence of looping.
\Sref{dl} shows how to extend the simulation to study looping, defines the looping $J$ factor, and gives results on $J$ as a function of loop length, both with and without the effect of the tethered
bead and surface, and for both the closed (V-shaped) and hypothesized open conformation of the lac repressor tetramer.
\Sref{elbe} gives a more refined measure of bead motion, the probability distribution of the bead excursions.
\Sref{rtor} discusses the relation between our work and earlier theoretical papers, and finally \sref{c} gives general discussion.

The Supplementary Information has its own table of contents; it contains information more directly related to the experimental data, details of our Monte Carlo algorithm, and some additional
calculations in our model. For example, we checked our work by calculating cyclization $J$ factors and comparing to the classic Shimada--Yamakawa result.

\section{Survey of experiments on looping\pnlabel{s:stel}}
Experimental measurements of DNA loop formation have fallen into four main classes. Readers familiar with the experiments may wish to skip directly to \sref{etu}.

\paragraph{Cyclization}In these \textit{in vitro} experiments, many identical, linear DNA constructs are prepared with overhanging, complementary ends. Ligase enzyme captures transient states in which either
two ends of the same DNA join, forming a ring, or else ends of two different DNAs join, forming a dimer. Under suitable conditions the ratio of rings to dimers after the reaction runs to completion
gives information about the equilibrium populations of those paired states, and hence about loop formation (e.g.,{} \cite{Shore81,Shore83a,kahn92a,clou05a,du05a}). Unfortunately, the interpretation of these
experiments is complicated by the role of the large, complex
ligase enzyme, the need to be in a very specific kinetic regime, and so on \cite{du05b}. Moreover, the process of interest to gene regulation is looping, which is geometrically quite different from cyclization.

\paragraph{{\itshape In vivo} repression}Other experiments measured the output of an operon as its controlling promoter was switched by a repressor (e.g.,
\cite{Muller1996,vila05a,beck05a,law93a,beck07a}); theory then connects those results to
looping $J$ factor values (or looping free energy changes) \cite{bint05b,bint05c,Saiz2005,zhan06b}. Although the experiments showed that
short loops form surprisingly easily, their quantitative interpretation is obscured by uncertainties due to the complex world inside a living cell, for example, supercoiling and
the many other DNA-binding proteins (such as HU, H-NS and IHF) present in cells. 

\paragraph{Magnetic tweezer}To introduce supercoiling in an \textit{in vitro} preparation, some experiments manipulate the DNA using a magnetic bead in a trap. Some earlier implementations unavoidably also
introduced extensional stress on the DNA  \cite{lia03a}; however, recent work has overcome this limitation \cite{norm08a}.

\paragraph{Tethered particle}In the present work we study TPM experiments \cite{scha91a}, which can report directly on looping state under controlled, \textit{in vitro} conditions.  Recent work on looping via TPM includes
\Nrrefs{zurl06a,broe06a,vanz06a,beau07a,beau07b,%
han?cLength,han?aCalib}. TPM experiments do require significant analysis to
determine looping state from bead motion, but techniques such as dead-time correction \cite{vanz06a} and Hidden Markov modeling \cite{beau07a,beau07b} now exist to handle this. Like cyclization, the TPM
experiments we studied have the biologically unrealistic (but theoretically convenient) feature that the supercoiling stress applied externally to the loop is zero.
({For a theoretical approach to looping with supercoiling see e.g.,{} \Nrref{puro06a}.})

Additional advantages of TPM include the fact that it does not involve fluorescence, and so is not subject to bleaching; thus an experiment can generate an unlimited data sample simply by tracking a
bead for a long time. Moreover, the DNA is in solution, and minimally affected by the distant bead.
Some implementations of TPM do not track individual trajectories, instead observing the blurred
average image of each bead \cite{finz95a,wong07a}; this article will focus on particle-tracking implementations (see, e.g.,  \Nrrefs{poug04a,nels06a}). Other
experimental aspects, including the attachment of the DNA of interest to the mobile bead at one end and the immobile surface at the other, are discussed in the original articles cited above.

TPM experiments also offer the ability to separate the overall probability of looping, at least partially, into the contributions of individual loop types (see \sref{elbe}). This additional degree of resolution allows
more detailed comparison with experiment than is possible when we observe only the level of gene repression.
Finally, TPM and other \textit{in vitro} methods also present the opportunity to dissect the experimentally observed looping probability into separate numerical values for the looping $J$ factor and the binding constant, via a
titration curve (discussed in \aref{edlJf}). In contrast, some \textit{in vivo} methods must obtain a value for the binding constant from a single data point (repression with auxiliary operator deleted), and
moreover must rely on the accuracy of an estimate for the effective repressor concentration in the cell \cite{Saiz2005}.

\section{Elasticity theory used in this paper\pnlabel{s:etu}}
This section derives the elastic model of DNA to be used in this paper. \Sref{emtp} first obtains the elasticity matrix up to an overall constant from structural information; then \sref{pl} fixes the
constant by requiring a particular value for the persistence length. Our simulation method involves matrix exponentiation, and may be simpler than other methods sometimes used in the literature.

\subsection{General framework\pnlabel{s:etgf}}
The physical model of DNA as a uniform, isotropic, slender, linearly-elastic rod \cite{volobook} has proven to give an adequate description of DNA mechanics for some purposes, notably for computing the force--extension
relation of long DNA \cite{PhilBook,bouc99a}. However, this simple model is not obviously appropriate for describing the formation of structures involving DNA loops of length comparable to a
helical repeat ($\ell_{\rm helix}=3.5\,\nmunit$). For example, in this paper we are interested in loops as short as 9 times the helical repeat length. On length scales comparable to $\ell_{\rm
helix}$, the bend stiffness anisotropy of the molecule certainly becomes
significant, as well as elastic cross-coupling between bend and twist \cite{mark94a,esla07a}. \Sref{emtp} below spells out the details of the elasticity theory we will use. (\Aref{ciem} explores the importance of
including the anisotropy by studying an alternative model.)

\ifigsize{FrameNotation}{Basepair geometry~\Nrref{mark94a}. The rectangle represents a DNA basepair. The red and blue dots are the phosphate
backbones. The circle is the outer envelope of the double helix, 2$\,\nmunit$ in diameter. We set up an orthonormal frame (\emph{left}) where $\Nbz $ is out of the page, $\Nbx $ points to the major groove,
and $\Nby $ completes the triad and points towards the $5^{\prime} \rightarrow 3^{\prime}$ strand (red) as defined by the positive $\Nbz$ direction. ``Positive roll'' is then defined as a positive
rotation about $\Nby $ as we pass from this basepair to the one on top of it (=``bend into the major groove''). Similarly ``tilt'' is rotation about $\Nbx $, and ``twist'' is excess rotation about
$\Nbz $ (in addition to the natural helical twist). For our purposes, a DNA chain conformation is a sequence of such frames. Graphically we represent it in \frefs{cartoon} and \ref{f:goodbad}
as a chain of double-helical segments, as shown on the \textit{right}.
}{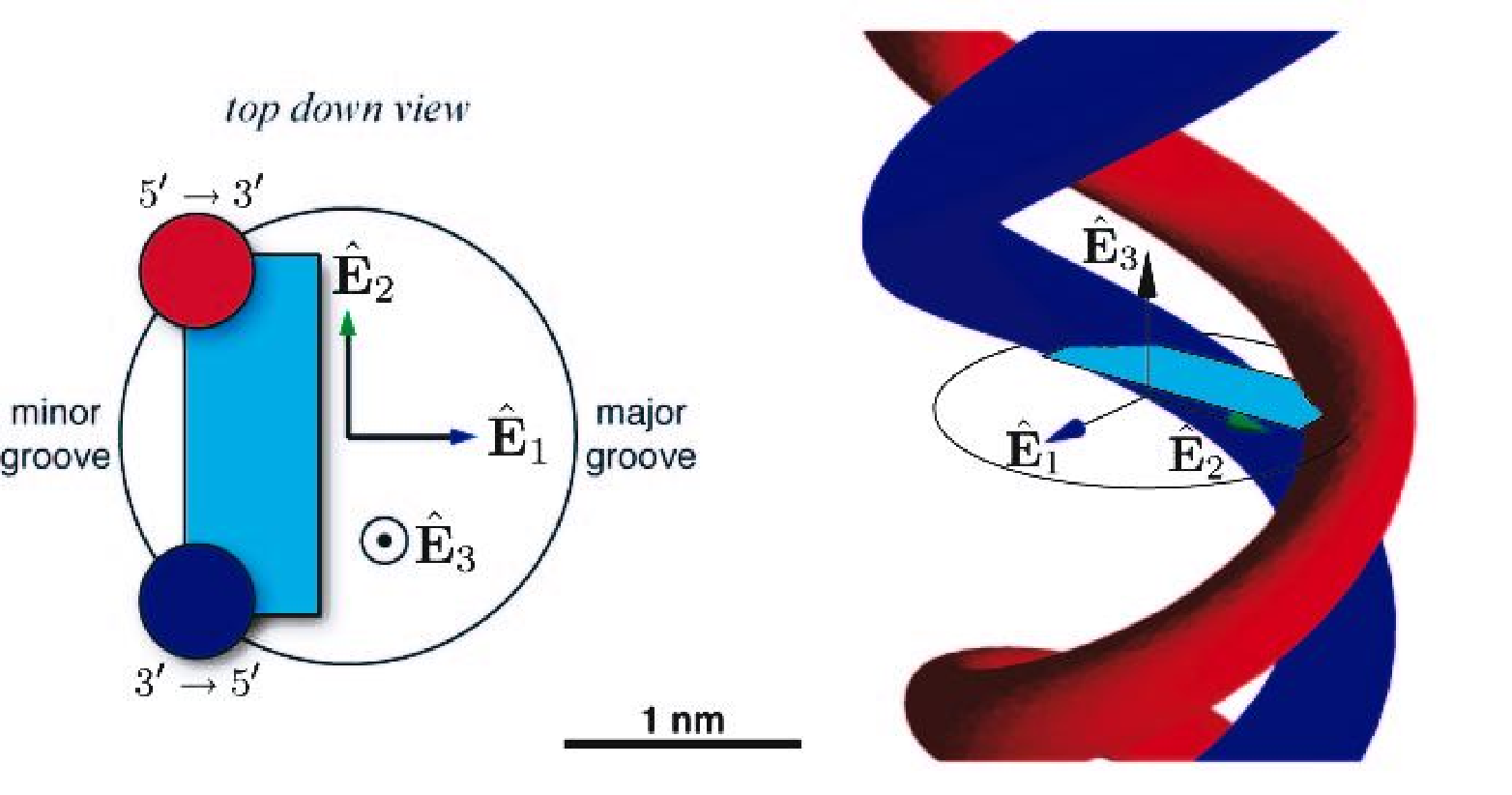}{4.5}
In other respects, our elastic model will be standard. We assume that the unstressed state of DNA may be regarded as a stack of plates (``segments''), each with thickness $\ell_{0}$ and each with a chosen reference
point and an inscribed coordinate frame at that point (\fref{FrameNotation}). Each plate is shifted a distance $\ell_0$ along its $\Nbz$-axis relative to its predecessor, and also rotated
by $2\pi \ell_0 /(\ell_{\rm helix})$ about the same axis. Next we need to quantify the elastic energy cost for a deviation from this unstressed state.

We restrict attention to a harmonic elasticity model, that is, we assume that the elastic energy at each junction is a quadratic function of bend
and excess twist, neglecting the  possibility of elastic breakdown at high strain \cite{yan04a,wigg05a,clou05a}. We do this because ultimately we are interested in \textit{testing} the harmonic model, by
confronting its predictions with experiment, and also because there is not yet a unique candidate for the detailed, three-dimensional form of an effective nonlinear elastic function.

We neglect stretch elasticity of the segments because there is no externally applied stretching force in TPM experiments (and any entropic stretching force is insignificant in this context
\cite{sega06a}). Thus the displacement of each segment is always $\ell_0$; the
``pose'' (position and orientation) of each segment relative to its predecessor is completely specified by the angular orientation. For simplicity, we also neglect
the sequence-dependence of DNA elasticity, so our results will apply only to random-sequence DNA constructs; all our comparisons to experiments will involve DNA of this type.
Because we are making a finite-element approximation to a continuum elasticity model, we
have some freedom in choosing the contour length $\ell_0$ of each segment, as long as it is much shorter than the persistence length, about 150~basepairs. To speed up calculations, we have chosen a segment length
corresponding to one fifth of a helical repeat (about 2~basepairs). Making our segments commensurate with
the helical repeat also has the advantage of showing clearly any helical phasing effects, i.e., modulation of looping with period equal to $\ell_{\rm helix}$.

Let $\Delta\theta_i$ be the excess rotation angles (beyond the natural twist) from one segment to the next and let
$\Omega_i=\Delta\theta_i/\ell_{0}$ denote the corresponding strain rates per unit contour length, where $i=1,2,3$ correspond to tilt, roll, and twist (see \fref{FrameNotation}).
We will define the elastic deformation free energy per unit contour length as
\begin{equation}\Nen\equiv(\half\kbt)\Omega^{t}\Qmat\Omega
,\label{e:caE}\end{equation}
so the stiffness matrix $\Qmat$ has units of length and is independent of the choice of segment length $\ell_0$. (The compliance matrix is then $\Qmat\inv$.) In the traditional wormlike chain model
$\Qmat$ is diagonal, with the bend and twist persistence lengths on the diagonal. We next propose a more realistic choice for this matrix.

\subsection{Relative elastic constants\pnlabel{s:emtp}}
To get values for the elements of $\Qmat$, we first note that (neglecting sequence-dependence) B-form DNA has a symmetry under $180^{\circ}$ rotation about any line perpendicular to its long axis and passing
through its major groove. (Such a line is labeled $\Nbx $ in \fref{FrameNotation}.) This symmetry forbids any harmonic-elasticity coupling between twist and tilt (that is, between small rotations about
$\Nbz $ and $\Nbx$ in the figure), and also between tilt and roll \cite{mark94a}. Thus the symmetric $3\times3$ matrix $\Qmat$ has only four independent nonzero entries \cite{mark94a,Moroz1998}.

Next, we adapt a strategy used by W.~Olson and coworkers \cite{olson98a}, who examined crystal structures of many DNA oligomers and of DNA--protein
complexes.  They then supposed that each basepair is subjected to random external forces (e.g.,\ crystal forces), the same for every type of basepair junction, analogous to the random forces in
thermal equilibrium but of an unknown overall magnitude. The observed deformations of basepairs in this imagined random external force tell us about the elastic compliances for deformation of each
basepair type, and in particular the covariances of deformations give the off-diagonal terms. Finally, we adjust the overall scale of the resulting elastic-energy matrix to obtain the desired persistence length of DNA in the
buffer conditions appropriate to the TPM experiments of interest.

The method outlined above, although rough, nevertheless captures the basic structure of DNA elasticity while preserving the required overall persistence length.
To carry it out,
we took the published covariance matrices for the $\Delta\theta_{i}$ of various basepair steps \cite{olson98a} and averaged them to obtain an elastic compliance matrix. We inverted this
matrix and  observed that indeed
the $(12),\ (13),\ (21),\ (31)$ entries of $\Qmat$ were much smaller than the others; we subsequently set them to be exactly zero. These steps yielded the entries of $\Qmat$, up to an overall
scale factor, as
\begin{equation}\Qmat=\Npconst\times
\left[
\begin{array}{lll}
0.084 & 0&0 \\
0 & 0.046&0.016\\
0&0.016&0.047\\
\end{array}
\right].
\label{e:uptoconst}\end{equation}
The overall constant $\Npconst$ has units of length; it will be specified in \sref{pl}.

The expected anisotropy is evident in the form of the matrix: The tilt eigenvalue (0.084) is much larger than the smaller of the two remaining eigenvalues (0.030).
Note that the near-degeneracy of the last two diagonal elements means that the eigenvectors are strongly mixed: The smaller eigenvalue corresponds
to a mixed deformation, with positive roll and negative twist. Thus, bending the DNA tends to untwist it \cite{olson98a}. Note, too, that the numerical values of the diagonal entries are not a good
guide to the relative  actual bend stiffnesses, because the eigenvalues of the $2\times2$ submatrix may be quite different from its diagonal entries.

\subsection{Specification of overall scale factor\pnlabel{s:pl}}
The persistence length $\xi$ of a polymer is defined by the falloff in correlation between the long-axis directions of nearby elements when the polymer is free (no external forces). Thus
$\exv{\Nbz (s)\cdot\Nbz (s+t)}\to\ex{-|t|/\xi}$ at large $t$, where $s,t$ are
contour lengths \cite{PhilBook}. We now discuss how to compute $\xi$ for an elastic matrix of the form \eref{uptoconst}, as a function of the unknown parameter $\Npconst$ that sets the strength of
$\Qmat$; demanding a particular value of $\xi$ will then fix the value of $\Npconst$. (A similar discussion recently appeared in \cite{moro?a}.)

To compute $\xi$ given a choice of $\Npconst$, we first generate a string of random rotation matrices, each representing relative rotations of one segment relative to its predecessor. These matrices are drawn from a distribution
that is centered on the identity matrix and weighted by the Boltzmann factor $\ex{-\Nen\ell_0/\kbt}$. More explicitly, we choose a value of
$\ell_0$, then diagonalize the matrix $\Qmat/\ell_0$, writing it as $\Tmat^t\Dmat\Tmat$ for an orthogonal matrix $\Tmat$. We then  use the diagonal entries of $\Dmat$ as inverse variances for
three Gaussian random variables $\{\Psi_i\}$, and let $\Delta\theta=\Tmat^t\Psi$, obtaining three random variables $\Delta\theta_i$ with the desired statistical properties.
We convert these random angles into a rotation
matrix by computing the matrix exponential $\exp\bigl(\sum_{i=1}^3\Delta\theta_i\Jmat_i
\bigr)$, where $\Jmat_i$ are the rotation generator matrices. (For example, $\Jmat_3=\threethree{0\, -1 \ 0}{1\ \ 0 \ \,0}{0\ \ 0 \ \,0}$.) Finally, we multiply the resulting rotation on the left by the
natural, unstressed DNA rotation $\exp\bigl((2\pi\ell_0/\ell_{\rm helix})\Jmat_3\bigr)$, obtaining $\Rmat(1)$, then repeat all these steps to make a long string of matrices $\Rmat(1),\,\Rmat(2),\cdots$.

Next, we step through the matrix string, cumulatively applying each rotation $\Rmat(k)$ in turn to an initial orientation to obtain the orientations of successive basepairs from a standard orientation for the
first one. That is, let the frame vector at arclength position $s$ be $\Nbgeneric a(s)$. We express it in components using the fixed lab frame as $[\Nbgeneric a(s)]_i=\hmat_{ia}(k)$, $i=1,2,3$,
where $s=k\ell_0$, $\hmat_{ia}(0)=\delta_{ia}$, and $\hmat(k+1)=\hmat(k)\Rmat(k)$. Finally we average the quantity $\exv{\Nbz (s)\cdot\Nbz (s+t)}$ over the generated chains, average over $s$ for various fixed $t$, confirm the
exponential decay in $t$, and extract the decay length $\xi$.

In solvent conditions used for TPM by Han et al.\ \cite{han?aCalib,han?cLength%
}, the persistence length has been previously measured by other means to be around $44\,\nmunit$
\cite{stri98a,wang97a}; see also \sref{fsr}, where we show that this value is consistent with TPM calibration data. Applying the above procedure to \eref{uptoconst} and requiring $\xi=44\,\nmunit$ fixes $\Npconst$: We then have
\begin{equation}\Qmat=\left[
\begin{array}{lll}
67\,\nmunit&0&0\\
0&37\,\nmunit&13.0\,\nmunit\\
0&13.0\,\nmunit&37\,\nmunit
\end{array}
\right].
\label{e:Qmat}\end{equation}
\Eref{Qmat} is the form suitable for angles $\Delta\theta$ expressed in radians; for angles in degrees the matrix should be multiplied by $\left({\pi}/{180}\right)^2$.

\section{Calculation of TPM distributions without looping\pnlabel{s:ctdwl}}
This section introduces our Monte Carlo method, and
gives a crucial check that theory and experiment are both working properly,
by showing to what extent we can accurately predict the excursion of the tethered bead in the absence of looping. Some details relevant to experimental data (blur correction and finite-sample
effects) are relegated to the Supplement.

We begin our analysis by predicting the motion of a tethered particle in terms of the tether length and bead size, both of which were systematically varied in the experiments of
\Nrref{han?aCalib}. Besides being a basic polymer science question, such {\it a priori}
knowledge of, say, the root-mean-square bead excursion for simple tethers sets the stage for our calculations involving looped tethers in \sref{elbe}. More generally, in other kinds of experiments the
tether length may be
changing in time, in a way that we would like to measure, as a processive enzyme walks along  DNA or RNA \cite{vanz03a}, or as proteins bind to the DNA, etc. Finally,
by comparing theory to experiment, we gain confidence both that the experiment is working as desired and that our underlying assumptions about the polymer mechanics, bead--wall interactions, and so
on, are adequate. 

Although the end-end distribution of a semiflexible polymer such as DNA is a classical problem in polymer physics, the present problem differs from that one in several respects. For example, the
DNA is not isolated, but instead is attached to a planar surface, and hence experiences an effective entropic stretching force due to the steric exclusion from half of
space; a similar effective repulsion exists between the DNA and the large bead.  More important than these effects, however, is the steric exclusion of the bead from the wall.
\NRref{sega06a} argued that the effect of this exclusion would be to create an entropic stretching force on the DNA.

Additional subtleties of the problem include the fact that the polymer itself has two additional length scales in addition to the bead radius, namely its persistence length $\xi$ and total length $L$,
and the fact that we do not observe the polymer endpoint, but rather the center of the attached bead. Some of these effects have been studied analytically for the case with applied
stretching force (e.g.,{} \cite{seol07a}), but
for \textit{zero} applied stretching force the steric constraints, not fully treatable in that formalism, become important. For this reason, \Nrrefs{sega06a,nels06a} developed a Monte Carlo calculation
method.\Nfootnote{fnu}{\NRrefs{bouc06a,bouc07a} studied related spatial-constraint effects in an
analytical formalism; the present paper gives a numerical approach.} A similar method was independently used for a study of DNA cyclization by Czapla et al.~\cite{czap06a}, who call it ``Gaussian sampling.''
Here we generalize that method to use the elasticity theory described in \sref{etu}.
We also extend our earlier work by computing the dependence of the RMS bead excursion on both tether length and bead size, and comparing to
experimental data in which both were systematically varied.

\subsection{Gaussian sampling\pnlabel{s:gs}}
The Gaussian sampling approach is not a  Markov-chain algorithm; each chain is generated independently of all the others, in the Boltzmann distribution associated with the elastic energy function. What makes this
approach
feasible is that the elastic energy functions of each junction between links are all independent (because we assume that there is no cooperativity between basepairs separated by more than our segment
length $\ell_0$). Thus, the random bends between links are also independent; we generate a chain by creating a string of rotation matrices each generated as described in \sref{pl}. To implement the steric
constraints, we next suppose additional energy terms of hard-wall type (i.e.\ either zero or infinity). Although it is an approximation to real mesoscopic force functions,
the hard-wall approximation is reasonable in the high salt conditions studied in typical experiments. Together with the approximate representation of a real microscope slide as a perfect plane (a
``wall''), it has proven successful in our earlier work \cite{nels06a}.

The constraint energy terms set the probability of the sterically-forbidden chains to zero. In practice, then, we generate many chains, find each chain segment's spatial position (and that of the
bead) by following the $\Nbz$-axis of each orientation triad, and discard the chain if any steric constraint is violated. All our thermodynamic averages are then taken over the remaining (``allowed'')
chains. For short tethers, many chains will be discarded, but as long as the fraction of ``allowed'' chains is not too small the procedure is tractable.

We treat the biotin and digoxigenin linkages attaching the DNA to bead and wall as freely flexible pivots, and so the orientation of the first chain segment, and that of the bead relative to the
last segment, are taken to be uniformly distributed in the half-spaces allowed by the respective surfaces. This approach has previously been successful in explaining experimental results
\cite{sega06a,nels06a,li06a,seol07a}. That is, the initial chain segment's orientation is a uniformly distributed random rotation subject to the half-space constraint; subsequent segments are then
determined by successive matrix multiplication by the rotations distributed as in \sref{pl}; the final vector $\Nmvec$ describing the bead orientation relative to its attachment point (black arrow
in \frefn{cartoon}a) is again taken to be uniformly distributed in the half-space defined by the final chain segment.

The steric constraints we implemented were \textit{(i)}~chain--wall, \textit{(ii)}~chain--bead, and \textit{(iii)}~bead--wall exclusion. For the short DNA tethers we consider, chain--chain
excluded volume is not expected to be a significant effect (although it would be important if supercoiling stress were  applied to the bead \cite{Moroz1998}).

We can see the trends in the data more clearly if we reduce the distribution of bead position to the root-mean-square excursion $\Nrrms\equiv\sqrt{\exv{\rho^2}}$, a quantity often used in experiments to characterize
tethered particle motion. A closely related quantity is the finite-sample RMS excursion, for example $\Nrrmsfour\equiv\sqrt{\exv{\rho^2}_{\rm 4\, s}}$. Here the expectation value is limited to a sample consisting of
$(4\,\sunit)/(0.03\,\sunit)$ consecutive video frames at a frame rate of $1/(0.03\, \sunit)$. Note that whereas $\Nrrms$ is a single number for each bead-tether combination, in contrast $\Nrrmsfour$
has a probability distribution.
One of our goals in the remainder of this paper is to predict $\Nrrms$ (in this section), or the distribution of $\Nrrmsfour$ (in \sref{elbe}), as functions of bead size, tether length, and tether looping state.

\subsection{Calibration curve results\pnlabel{s:fsr}}
\ifigsize{calibrplot}{Theoretical prediction of equilibrium bead excursion. \textit{Dots:} Experimental values for RMS excursion of bead center, $\Nrrmst$, for random-sequence DNA and three different
bead sizes: Top to bottom, $\NRbead =485,\ 245,$ and $100\,\nmunit$. (Data from \Nrref{han?aCalib}.) The sampling times were $t=20,\ 10$, and $5\,\sunit$ respectively. For these rather long times the
finite-sample correction is negligible; nevertheless, we included this correction (via a method given in
\aref{sse}). Each dot represents $20$--$200$ different observed beads with the given tether length. Dots and their error bars were computed by the method
described in \fref{blurPhil}. \textit{Curves:} Theoretically predicted RMS motion, corrected for the blurring effect of finite shutter time. For each of the three bead sizes studied, two curves are
shown. From top to bottom, each pair of curves assumes persistence length values $\xi=47$ and $39\,\nmunit$, respectively. There are \textit{no fit parameters;} the theoretical model uses
values for bead diameter given by the manufacturer's specification.
}{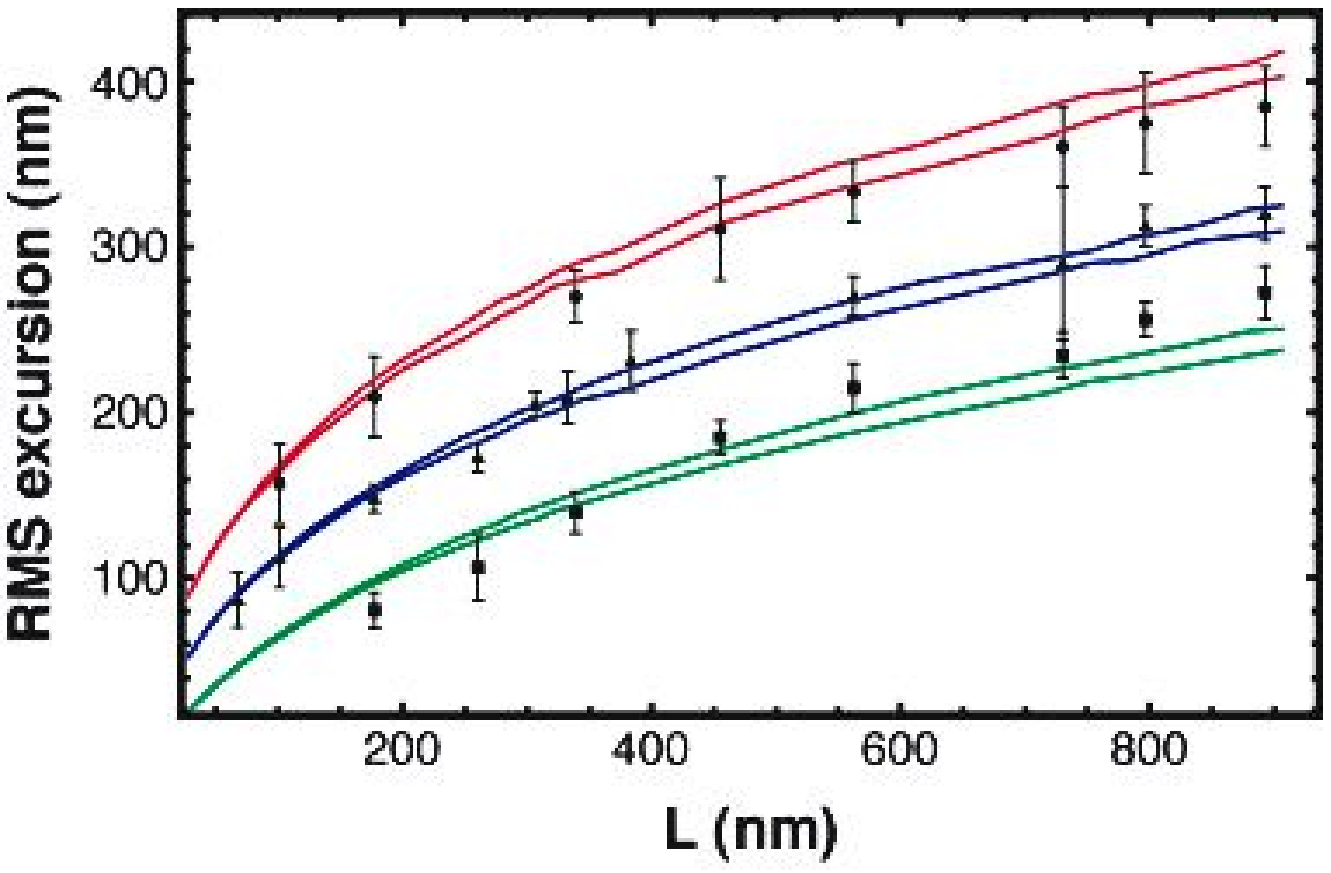}{4.5}
\Sref{gs} explained how, given values of $L$, $\NRbead$, and $\xi$, we generate many chain/bead configurations. From these configurations, we can in principle compute quantities like $\Nrrms$. (An
additional correction, to account for finite camera shutter speed, is explained in \aref{bc}.)
We compute $\Nrrmst$ in this way and compare it to the experiments of Han et al.\ \cite{han?aCalib}. We took $L$ to be $0.34\,\nmunit$ times the number of basepairs
in each construct, and accepted the manufacturer's specifications of $\NRbead$ for beads of three different sizes, leaving us with just one remaining parameter, the persistence length $\xi$.
The finite sampling times used in the experiment had an insignificant effect (data not shown), but nevertheless we included this aspect of the experiment (see \aref{sse}) for consistency with our later study of the
probability density function of bead excursion in \sref{elbe}. In that context, the finite sampling time is important.

DNA stretching experiments using high-salt buffer similar to that used in the TPM experiments we study obtained a persistence length of $\xi=45\,\nmunit$ \cite{stri98a}, or
$43\,\nmunit$ \cite{wang97a}. When we turn to TPM, \fref{calibrplot} shows that indeed taking $\xi$ in the range $39$--$47\,\nmunit$
reproduces the trends of the data fairly well with \textit{no fitting}, even though this is a very different class of experiment from stretching.
({Previous work came to a similar conclusion \cite{nels06a}, although it considered only a single bead size.})
The curve with bead size $245\,\nmunit$ is particularly well predicted; all TPM data appearing in the rest of this paper were taken with this value of $\NRbead$.
Throughout the rest of this paper we will use the value $\xi=44\,\nmunit$.

\section{DNA looping\pnlabel{s:dl}}
This section attempts to distill loop formation into a mathematical problem, the calculation of a quantity called the ``looping $J$ factor'' (\srefs{ltsls1}--\ref{s:lJf1}; some geometrical details
about the looping synapse are deferred to \aref{ltsls2}). \Aref{edlJf} in the Supplement explains how we extracted $J$ from experimental data.
Next, \sref{clJf} describes the calculation of $J$ (more details are in \arefs{acg}--\ref{s:MCa}) and \sref{Jfr} compares to experiment. For loops of length near 300$\,\bpunit$, our absolute prediction for $J$ agrees with
the preliminary experimental data now available to within about a factor of 3; equivalently the corresponding looping free energies agree to within about $1\,\kbt$. However, the hypotheses embodied
in our model cannot explain the observed $J$ factor for short loops, near 95$\,\bpunit$ between operators. We will argue that the hypothesis of an alternate ``open'' LacI conformation is not sufficient to
resolve this discrepancy.

\subsection{Geometric structure of the loop complex\pnlabel{s:ltsls1}}
\subsubsection{DNA construct\pnlabel{s:dc}}
The experiments of Han et al.~\cite{han?cLength} studied DNA looping for random-sequence DNA in two classes, forming ``long'' and ``short'' loops.
({They also studied special sequences \cite{han?bSeque}, which we do not discuss in the present paper.})
Both ``long'' and ``short'' loop DNA constructs had the general form
\begin{equation}\mbox{\small\tt wall-($N_1\,\bpunit$)-($N_2\,\bpunit$)-($N_3\,\bpunit$)-($N_4\,\bpunit$)-($N_5\,\bpunit$)-bead }
.\label{e:construct}\end{equation}
The ``short'' constructs had $N_{2}=20\,$bp (the \NOid{} operator), $N_4=21\,$bp (the \NOone{} operator), and $N_1 = 144\,$bp, $N_3 = 89+\Nspacing\,$bp, $N_5 = 171\,$bp, where $\Nspacing$ is an integer equal to 0, 5, or 11.
The ``long'' constructs had $N_2=21\,$bp (\NOone), $N_4=20\,$bp (\NOid), and $N_1 = 427\,$bp, $N_3 = 300+\Nspacing\,$bp, $N_5 = 132 -\Nspacing\,$bp, where $\Nspacing$ is an integer between 0 and
10.\Nfootnote{}{Some of the actual constructs used in the experiment differed from the simple formula above by 1--2~basepairs
\cite{han?cLength}.}
For the purpose of labeling loop topologies,
we choose a conventional direction along the DNA that runs from \NOid{} to \NOone. Thus for the ``short'' constructs this direction runs from the wall to the bead, whereas for the ``long''
constructs it runs from bead to wall.

The artificial sequence $\NOid$ (``ideal operator'') binds DNA more strongly than the wild type \NOone. In fact, in the range of \Nlaci{} values we study, \NOid{} is essentially
always bound \cite{han?cLength}, and the looping transition consists of binding/unbinding of the  already-bound LacI to \NOone.

\subsubsection{DNA binding and its degeneracy\pnlabel{s:bdpg}}
\ifigsize{CylinderRepresentation}{Cartoon of the LacI tetramer (solid shapes) bound to two operator DNA segments (shown as wireframes). The tetramer consists of dimers D1 and D2, with
binding heads H1--H4. The wireframes show in detail the
dispositions of the operators relative to each other, as given in Protein Data Bank entry
{\tt 1LBG.pdb}. In the present work we summarize the entire structure by
the six orthonormal frames shown, which represent the \textit{entry/exit} and \emph{center} frames discussed in the main text and  \aref{ltsls2}.
The axes with blue, green, and black arrowheads represent $\Nbx$, $\Nby$, and $\Nbz$
respectively. These six frames were determined from the PDB file by the method described in \aref{ltsls2}.}{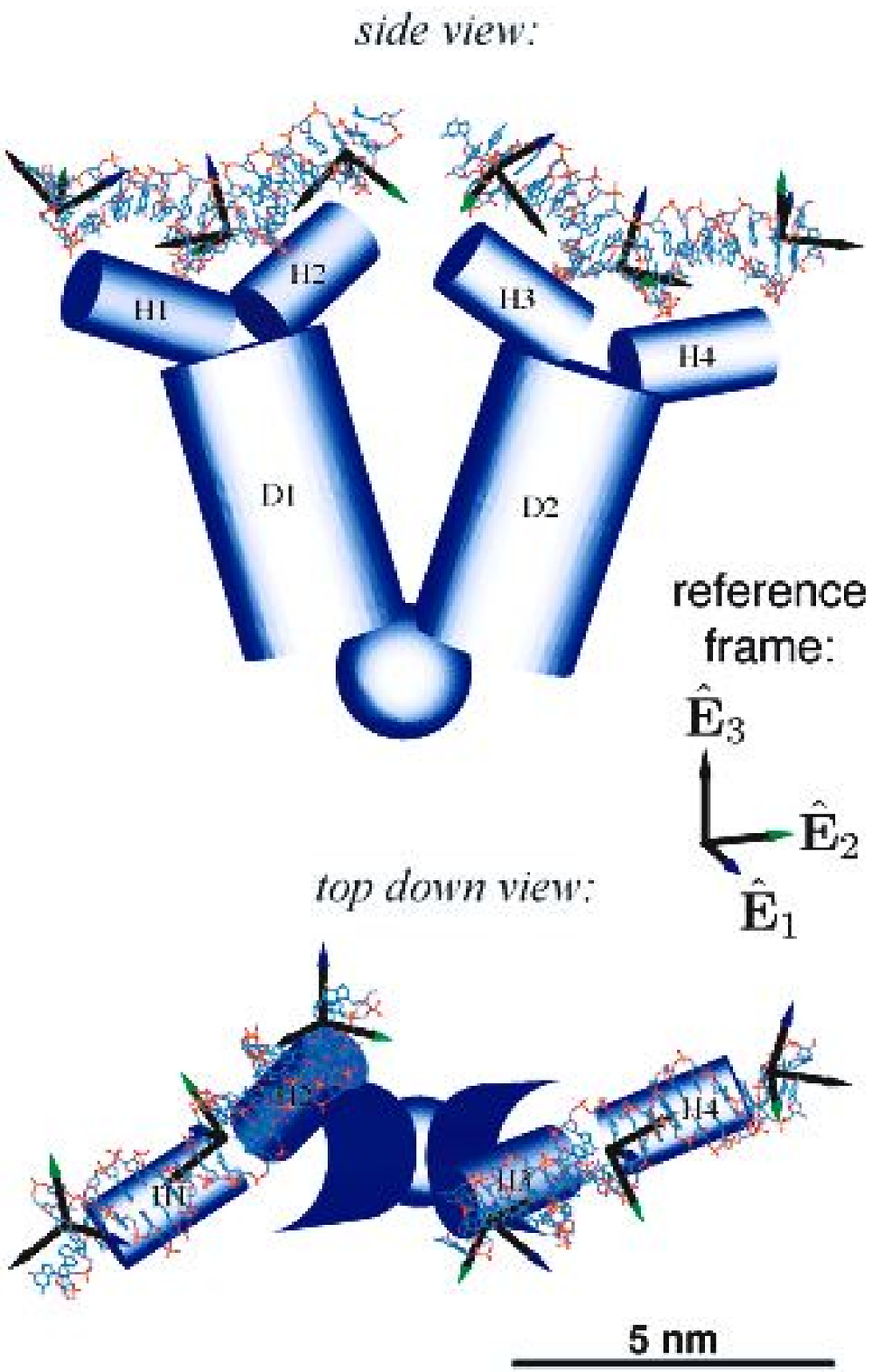}{2.7}
The LacI protein is a tetramer consisting of two identical dimers (D1, D2), each with two  heads (H1--H4) that  bind the DNA.\Nfootnote{fnt}{Lac repressor essentially always exists as
tetramers under the conditions of the experiments studied here \cite{Levandoski1996,Barry1999}.}
\Fref{CylinderRepresentation} shows a  cartoon, drawn to scale, based on the RCSB Protein Data Bank entry {\tt 1LBG.pdb} \cite{Lewis1996} (see also \cite{oehl90a,bell01a,bell01b}). Two segments of bound DNA
(operators of type \NOid) appear as well.
The cartoon is meant to portray the level of detail with which we treat the tetramer in our calculations: We regard the protein as a clamp holding the two bound operators rigid relative to each other. Thus, as soon as we specify the pose
(position and orientation) of the DNA bound to head H1 (say), we have also specified its exit from H2 as well as its entry and exit at H3 and H4.
\Fref{CylinderRepresentation}  shows six particular
poses, represented by orthonormal triads, associated to the \emph{entry}/\emph{exit} and \emph{center} basepairs. These are described in greater detail below and in \aref{ltsls2}. The axes are
color-coded; the blue, green, and black arrows correspond to the axis vectors $\Nbx$, $\Nby$, and $\Nbz$ in \fref{FrameNotation}.

Actually, each binding site has \textit{two} energetically equivalent binding orientations, due to a two-fold symmetry of the LacI dimer \cite{lewi05a}, so \fref{CylinderRepresentation} shows only one
of four possibilities. (The DNA sequence of the operator need not be a palindrome to have this degeneracy.) The symmetry operation on the DNA that relates these orientations is the same one described in
\sref{emtp}: $180^{\circ}$ rotation about the frame vector $\Nbx$ passing through the operator center and pointing to the major groove.

\ifigsize{molecules}{Four possible orientations of simulated looped chain (\textit{dashed lines}). Our convention is that the arrows run from \NOid{} to \NOone. Two binary variables describe the binding
orientation at the two operators as shown. If the chain exits \NOid{} at an inner headgroup (H2 or H3 in \fref{CylinderRepresentation}), we say $\beta=1$. If the chain enters \NOone{} at an inner
headgroup, we say $\alpha=1$. There are two
``parallel'' loop configurations (P1, P2), for which the entry and exit trajectories of the  chain have nearly parallel $\Nbz$ axes; likewise, there are two
``antiparallel'' loop configurations (A1, A2), for which the entry and exit trajectories of the  chain are nearly opposite. Configurations A1, A2 look equivalent under
the symmetry that reverses DNA direction and exchanges the two LacI dimers. However, this apparent degeneracy is broken
when we add the bead to one end, and the wall to the other.}{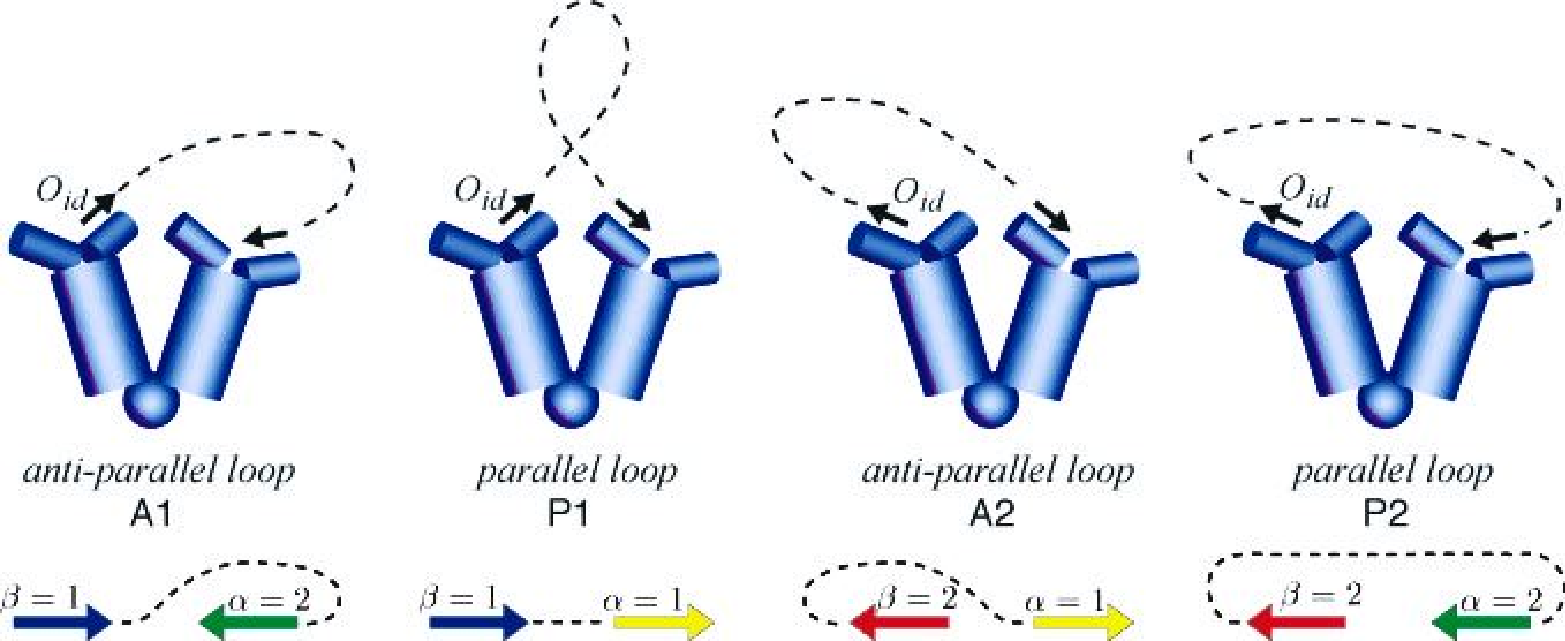}{6}

Referring to \eref{construct}, we will speak of the DNA as ``starting'' at the wall or bead, ``entering'' a binding site at one end of \NOid, ``exiting'' that binding site to the interoperator segment,
and (if looped) ``then entering'' the other site at \NOone{} and ``finally exiting'' to ``arrive'' at the bead or wall.
\Aref{ltsls2} describes our mathematical characterization of the geometry of LacI for the purposes of our simulation. Here we only note that because of an approximate twofold symmetry in the tetramer,
it is immaterial which dimer binds to \NOid. However, we do need to distinguish the two binding orientations at each site, because
they have inequivalent effects on the rest of the DNA.  We will distinguish them at \NOid{} by the label $\beta=1,2$. Similarly, we  introduce a label $\alpha=1,2$ denoting the binding
orientation at \NOone. \Fref{molecules} defines our conventions for these labels, which amount to specifying four topologically distinct classes of loops.\Nfootnote{fne}{Each of these classes in turn can
be further subdivided into distinct topoisomers. For example, we can take any of the loops shown in \fref{molecules}, detach the DNA from one binding head, twirl it about its axis by one full
revolution, then reattach it, resulting in a topologically distinct loop with the same values of $\alpha$ and $\beta$.
Because in experiments the topoisomer class of a loop can neither be observed nor controlled, however, we will not make any use of this subdivision in this
paper.}

\Fref{molecules} also
identifies each looping topology using names consistent with previous LacI looping studies \cite{Geanacopoulos2001,swig06b}. These topologies are grouped into two general categories
characterized by the relative orientation of the two bound operator sequences: parallel (P1, P2) or anti-parallel (A1, A2).

The dashed lines in \fref{molecules} represent the DNA loops and are added as visual aids; they are not results of our calculations. 

\subsection{The looping $J$ factor\pnlabel{s:lJf1}}
TPM (and some of the other experiments described in \sref{stel}) provides information about the fraction $\NPlooptot$ of time that a DNA tether spends in one of its looped conformations. Suppose that a
repressor tetramer is already bound to operator \NOid. Then we can regard
the looping transition as a combination of two subprocesses, namely \textit{(i)}~the occasional spontaneous bending of the DNA to bring  LacI and the other operator (\NOone) into proximity,
and \textit{(ii)}~binding of \NOone{} to LacI. The first of these processes will be characterized by a quantity called the ``looping $J$ factor'' below, whereas the second
is characterized by a chemical binding constant $\NKd$. The looping $J$ factor is the quantity of interest to us in this paper, as it is the one that we will subsequently attempt to predict
theoretically. It is a generalization of the classical $J$ factor from DNA cyclization  \cite{jaco50a,flor76a,mark82a}, which
can roughly be regarded as the concentration of one operator in the vicinity of the
other.
In this section we define $J$ mathematically; \aref{edlJf} describes how to obtain it from TPM data. ({\Aref{edlJf} will explain the relation between $J$ and the ``looping free energy change'' $\NDG$
discussed by other authors.}) \Sref{clJf} describes how we compute $J$ from our theoretical model, and \sref{TPMl} makes comparisons to available experimental results.

The overall dependence of looping on the length  of the intervening DNA between the operators can be qualitatively understood as reflecting  two competing phenomena. First, a short tether confines the
second operator into a small region about the first one, increasing the effective concentration. But if the required loop is too short, then forming it will entail a large bending elastic energy
cost, depressing the probability by a Boltzmann factor. For these reasons, the cyclization $J$ factor exhibits a peak at DNA length about 460$\,\bpunit$ \cite{shim84a}. Later work extended Shimada and
Yamakawa's calculation in many ways, using a variety of mathematical techniques
\cite{hage85a,leve86a,merl98a,podt99a,podt00a,zhan03a,rapa04a,spak04a,spak05a,ranj05a,yan05a,doua05a,zhan06a,zhan06b,wigg06a,spak06a,czap06a}; \sref{rtor} will comment on some of this work.

\ifigsize{goodbad}{Illustration of the notion of target pose with a representative looped chain from our simulations. The chain shown is considered to be  ``looped'' in the sense of \sref{lctc} because the 
center of its \NOone{} operator matches its target within a certain tolerance (shown not to scale by the \textit{blue-caged sphere}), and the orientation of the operator (\textit{small arrows} in the inset)  
aligns with the target orientation (\textit{large arrows} in the inset). DNA elasticity may favor thermal fluctuations that generate encounters with \NOone{} correctly oriented for binding (enhancing looping), 
or on the contrary, it may favor encounters incorrectly oriented for binding, depending in part on the number of basepairs between the two operators. \Frefs{IcosTangentShort}--\ref{f:IcosNormalShort} show this
phenomenon in our numerical results.}{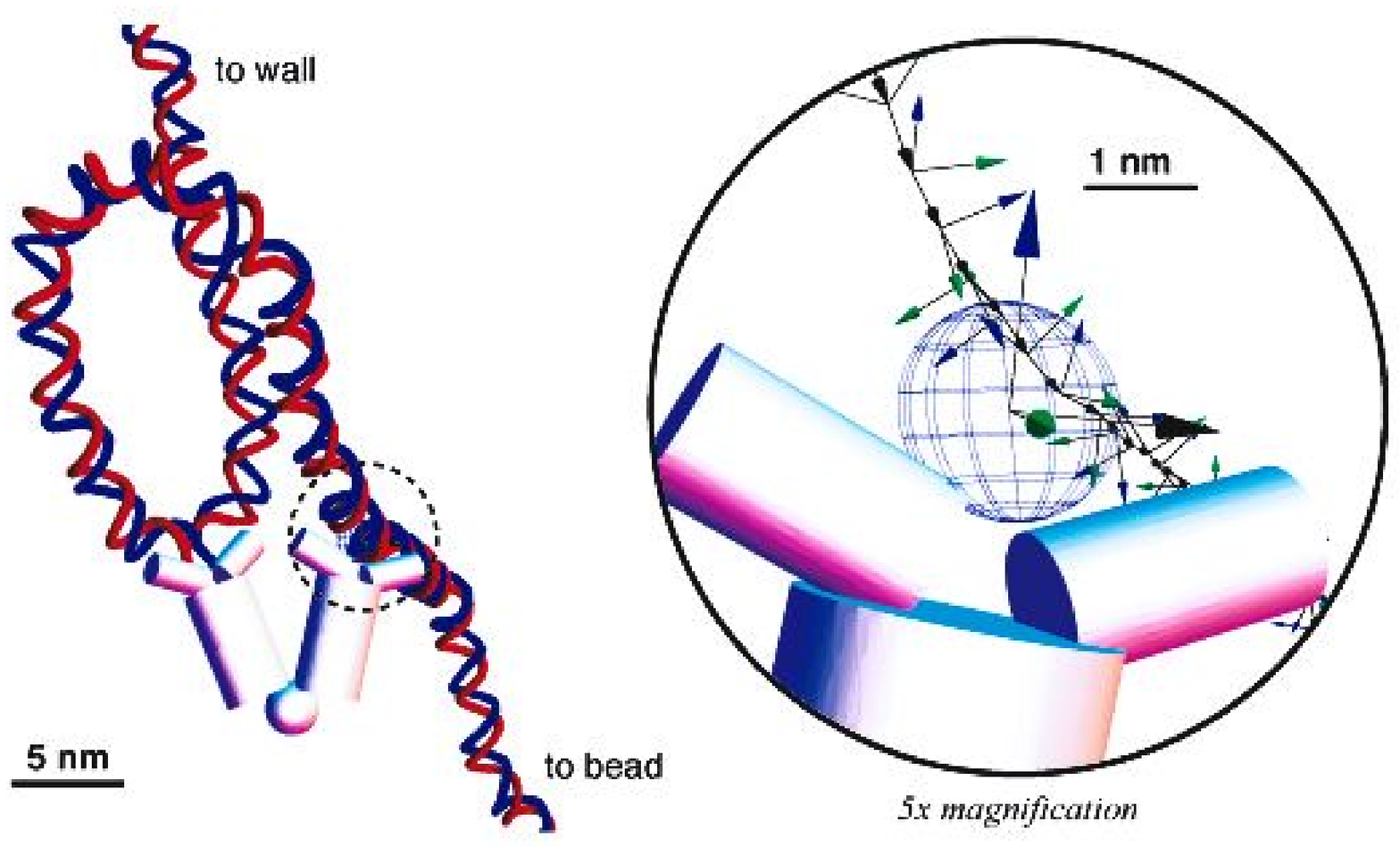}{4.5}

We now state the definition of the $J$ factor to be used in this paper, and introduce the closely related ``differential $J$ factor,'' which we call $\NJt$. As outlined above, we consider fluctuations of
the DNA chain conformation only, and ask how often operator \NOone's position and orientation fluctuate to coincide with a ``target'' representing the available binding site on a LacI tetramer already bound to
\NOid{} (see \fref{goodbad}). (A precise characterization of the target is given in \aref{ltsls2}.) A chain conformation is regarded as ``looped'' if the pose (position and orientation) of \NOone{}
matches the target to within certain tolerances. We express the spatial
tolerance as a small volume $\Ndv$ in space (with dimensions $(\mathrm{length})^3$), and the orientation tolerance as a small volume $\Ndo$ in the group of rotations, normalized so that the full group has
volume $8\pi^2$\cite{tung85a}. The total group volume may be regarded as solid angle $4\pi$ for the director $\Nbz$, times angular range $2\pi$ for the rotations of the frame about
$\Nbz$. Thus $\Ndo$ is dimensionless.

Our Gaussian sampling Monte Carlo code generates many DNA chain conformations in a Boltzmann distribution. If we suppose that a LacI tetramer is bound to \NOid{} with binding orientation $\beta$, then a
certain fraction of these chains are looped in the above sense with \NOone{} binding orientation $\alpha=1$; a different fraction are looped with  $\alpha=2$.
Clearly both of these fractions go to zero if we take the tolerances $\Ndv $ or
$\Ndo$ to be small, so we define  ``differential $J$ factors''  as
\begin{equation}\NJt^{(\beta)}_{\alpha}=\lim_{\Ndv,\Ndo \to0}\left(\mbox{fraction in looped conformation $\alpha$, given $\beta$}\right)/(\Ndv \Ndo)
.\label{e:dfJt}\end{equation}
It is convenient to introduce the abbreviations
\begin{equation}\NJt\tot\equiv{\textstyle{\frac14}}\sum_{\alpha,\beta}\NJt^{\,(\beta)}_{\alpha} \mbox{ \ and \ }\NJ=8\pi^2\NJt\tot
.\label{e:dfJnt}\end{equation}
Note that $\NJt$ and $J$ naturally carry the dimensions of concentration. Our justification for the conventions in \eref{dfJnt} is that $J$ defined in this way is a
generalization of the familiar cyclization $J$ factor \cite{jaco50a,flor76a,mark82a}. To see this, suppose that we consider a very long loop. Then whenever $\NOone$ wanders into its target volume, its orientation
will be isotropically distributed, and in particular all four of the $\NJt_{\alpha}^{(\beta)}$ are equal. If a LacI tetramer is bound to \NOid, then
the effective concentration $J$ of \NOone{} in the neighborhood of its other binding site (regardless of orientation) is related to the probabilities defined in \eref{dfJt} by (say)
$J=8\pi^2\NJt_1^{(1)}$. For arbitrary loop length (not necessarily long), we replace the last factor by its average, obtaining \eref{dfJnt}.\Nfootnote{fns}{For the case of cyclization there are no
labels $\alpha,\beta$ and no average; we then have $J=8\pi^2\NJt$, which with \eref{dfJt} agrees with the definition in \Nrref{czap06a}.}

$\NJt_{\alpha}^{(\beta)}$ depends on the position and orientation of the target; \sref{Jfr} will take these to be defined by the crystallographic structure of the repressor tetramer. But more generally, we
can regard $\NJt_{\alpha}^{(\beta)}$ as a function of \textit{arbitrary} target pose, which we will compute and display in \sref{odls}.

Although in principle TPM experiments can obtain the absolute magnitude of $J$, in practice the available experimental data are still sparse. Fortunately, the
\textit{ratio} of $J$ factors for two different situations is more readily obtainable than the absolute magnitude (see \aref{edlJf}). For this reason, we will sometimes report experimental values normalized to a mean
value $\bar J$, which we define as
\begin{equation}\NJb(\mathrm{long})=\mbox{mean of measured $J$ values over the range $300\le\NLl\le310\,$bp}
\ .\label{e:dfJb}\end{equation}

\subsection{Calculation of looping $J$ factor\pnlabel{s:clJf}}
\Arefs{acg}--\ref{s:MCa} describe how we generalized the Gaussian sampling Monte Carlo algorithm of \sref{gs} to handle looping. \Aref{cycl} describes how we checked our code, and our definitions such
as \erefs{dfJt}--\ref{e:dfJnt}), by calculating the \textit{cyclization} $J$ factor and comparing to the classic result of Shimada and Yamakawa.

\subsubsection{Orientation distribution of looped states\pnlabel{s:odls}}
\ifigsize{IcosTangentShort}{Distribution of the chain tangent vector for generated chains ending in the target volume (``hits,'' see \sref{lctc}) for the short construct tether. The possible
directions for $\Nbz$ at the center of \NOone{} have been divided into twenty bins and the observed probabilities to land in each bin are assigned colors. Each row of the figure shows an icosahedron
painted with the corresponding colors, from various viewpoints. The red faces correspond to the most populated bins; bluer faces correspond to lower hit densities. The four views represent clockwise
rotations of the viewpoint by $90^{\circ}$ about $\Nbz$ for the two binding orientations at \NOid. The reference coordinate frames at top represent the orientation of the \emph{exit} frame of $\NOid$.
Directions labeled P1, etc., refer to the target pose for the corresponding loop type, which does not in general agree with the most-populated bin. A total of about $7.5\ee{10}$ chains were generated, resulting in
1673 hits with $\beta=1$ and $12\,540$ hits with $\beta=2$.
}{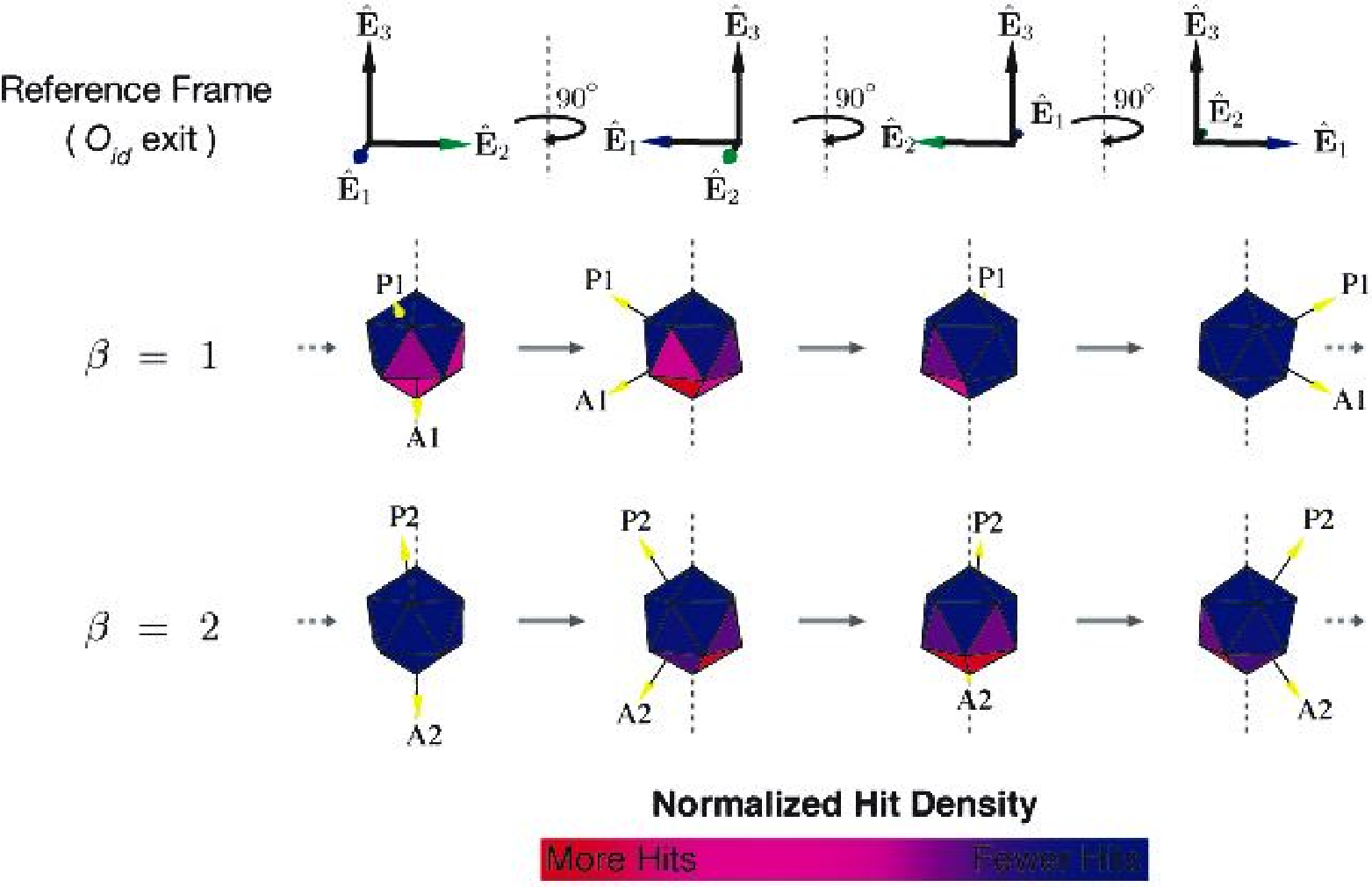}{5}
\ifigsize{IcosNormalShort}{Distribution of the normal vector $\Nbx$ for the short construct tether. Other conventions are similar to \fref{IcosTangentShort}, except the directions labeled P1, etc., correspond to the target normal
vector $\Nbx$ for the corresponding loop type.}{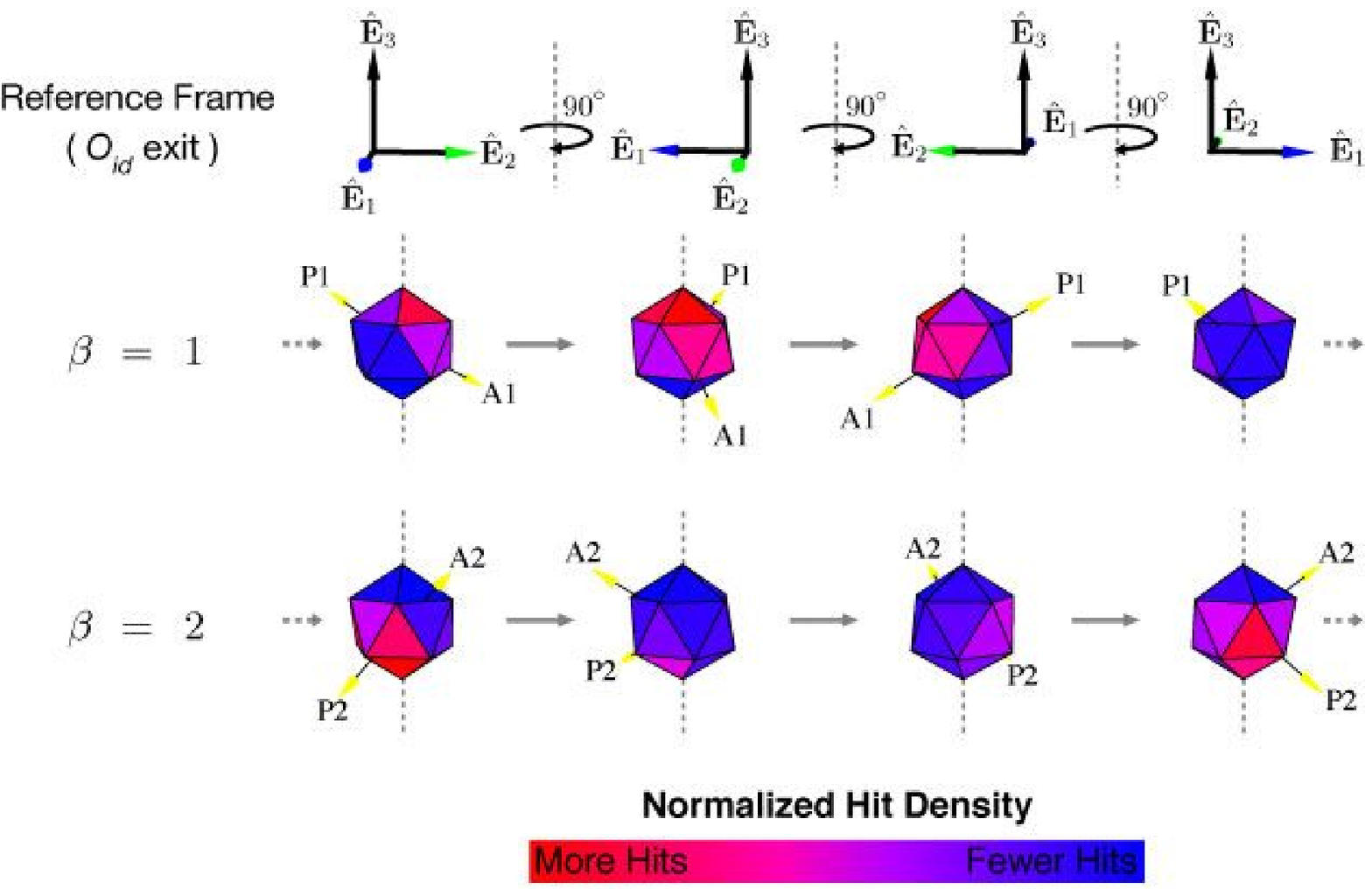}{5}
Each binding orientation of $\NOid$, with $\beta = 1, 2$, yields a characteristic distribution ${\cal C}_{\beta}$ of allowed chains, each with a particular pose for the center basepair of \NOone. Of
these, a small subset ${\cal C}^{*}_{\beta}$ will be ``hits,'' i.e.\ will have that center basepair inside its target volume for binding of the other site on LacI (see \fref{goodbad} inset). We are
ultimately interested in a smaller subset still, namely those chains ${\cal C}^{*}_{\beta\alpha}$ for which $\NOone$ is also in one of its two target orientations. First, however, it is instructive to
examine the \textit{distribution} of orientations for \NOone{} in ${\cal C}^{*}_{\beta}$. (The importance of this distribution was discussed long ago by Flory and coauthors \cite{flor76a}.)

For each ``hit'' configuration, we stored the orientation of the \NOone{} center segment relative to the exit segment of $\NOid$. \Frefs{IcosTangentShort} and \ref{f:IcosNormalShort} show the
distribution of the tangential ($\Nbz$) and normal vectors ($\Nbx$), respectively, for the ``short'' loop construct with loop length equal to $89\,$bp ($\Nspacing=0$ in the notation of \eref{construct}).
In these graphs we have taken the unit sphere and divided into 20 finite-solid-angle bins. The coloring shown on each face of the icosahedron represents the population of the corresponding angular bin.

\Frefs{IcosTangentShort} and \ref{f:IcosNormalShort} show that the orientation of hits is quite anisotropic, and not in general peaked in the target orientation for forming
any type of loop. These trends are characteristic of all loop lengths; however, the same plots of the ``long'' loop construct (not shown) reveal a broader, though still peaked,
distribution. The broadening of the distribution as the loop length increases is to be expected and is a natural consequence of the lability of long DNA loops. As the loop length
is increased one segment at a time, the distribution of the tangential vector evolves slowly, but the peak of normal vector distribution rotates with each added segment by about $2\pi\ell_0/\ell_{\rm helix}\approx
2 \pi /5$ radians (data not shown). This rotation of the normal
vector distribution with changes in loop length corresponds to the helical nature of DNA; as the peak rotates about the fixed target orientation, we get an approximately periodic modulation in the
$J$ factor called ``helical phasing'' \cite{Oehler1994}. A more
quantitative treatment of this behavior follows in \sref{Jfr}.

\subsubsection{Looping criteria and tolerance choices\pnlabel{s:lctc}}
Chains generated with the target segment located within the target volume $\Ndv$ (``hits'') pass the first constraint, the spatial tolerance check, as mentioned above (see also \fref{goodbad}). All results
correspond to a spatial tolerance of $\Ndv=(4\pi/3)(2\,\nmunit)^3$. Classification
of chains as looped or not is further dependent on an orientational constraint defined by $\Ndo$. We required that the tangent vector to the chain, $\Nbz$, at the center of \NOone{} lie within a cone
of angular radius $\pi / 4$ radians of the target direction. We also required that the major-groove direction at the center of
\NOone, $\Nbx$, projected to the 1-2 plane of the target orientation, must match the corresponding target frame vector to within $2\pi/5\,$radians. In other words, we checked whether the orientation of the major groove of the
generated chain's central operator segment matches its target orientation. If both of these conditions are met, the ``hit'' conformation is considered to be ``looped.'' The group volume corresponding
to these angular tolerances is thus
$\Ndo=2\pi(1-\cos\frac\pi4)(2 \times \frac{2\pi}{5})\approx4.63$, which is much smaller than the full group volume $8\pi^2$. After a chain is classified as looped or not,
we proceed as described in \aref{MCa}.

According to \eref{dfJt}, we are interested in a limit as the tolerances $\delta v,\ \delta\omega$ approach zero. In practice we must of course keep these quantities finite, but we checked that we
were reasonably close to the limiting behavior by checking two other choices of these tolerances: We cut the spatial tolerance in half, leaving the orientational tolerances the same, and we cut the
orientational tolerances in half, leaving the spatial tolerance at $(4\pi/3)(2\,\nmunit)^3$. We found that, although the magnitude of the phasing oscillations
increased slightly for each reduction of the tolerances, nevertheless in each case the qualitative effect on  the $J$ factor calculations (and also on the RMS probability distributions, \sref{elbe})
was minimal (data not shown).

We have chosen to report results of the larger tolerance for two reasons: First, the number of hits is proportional to the tolerance, so we obtain better statistics with larger tolerance; second,
larger tolerances may actually do a better job of representing the real experimental situation, specifically flexibility in the head regions of the Lac repressor, which we do not otherwise include.
Recent all-atom simulations suggest that this flexibility is substantial \cite{Villa2005}.

\subsection{$J$ factor results\pnlabel{s:Jfr}}
Before presenting results for looping in TPM experiments, we briefly describe a simpler warmup calculation. Then \sref{TPMl} describes a calculation that can be compared to TPM data, with
moderately good agreement; \sref{RMSr} shows a much more striking agreement of theory with another kind of TPM data.

\subsubsection{Pure looping\pnlabel{s:Jfpl}}
One can imagine an experiment involving a DNA construct with only the two operators and the basepairs between them, that is, no flanking segments joining the loop to a wall and a bead.
Here we present results on this form of the looping $J$ factor (``pure looping''). We will also plot our results alongside corresponding experimental numbers for \textit{in vivo} looping, even though the latter
correspond to rather different physical conditions.

\ifigsize{gDNA}{$J$ factor for pure looping, as a function of loop length $\NLl$ in basepairs. The vertical axis shows minus the natural logarithm of $J$ (measured in molar). (Some authors call
this quantity $\NDG/\kbt$; see \aref{SV}.) Thus, higher points on the curves indicate more difficult looping; the curve rises at the left because of the high elastic energy cost of a short loop.
The triangle at 460$\,\bpunit$ roughly corresponds to the minimum of the overall looping $J$ factor. \textit{Dots:} Our Monte Carlo results. Blue, red, green, and cyan represent the quantities $2\pi^2\NJt_{\alpha}^{(\beta)}$
corresponding to P1, P2, A1, and A2 loops, respectively. \textit{Curves:} Each set has been summarized by an interpolating function described in \sref{Jfpl}. \textit{Black curve:} The sum of the colored curves, that is,
the overall looping $J$ factor assuming that each looping topology is equally weighted (see \eref{dfJnt}).  \textit{Inset:} An enlarged portion of the graph for loop lengths of $\approx300$--$330\,\bpunit$.
}{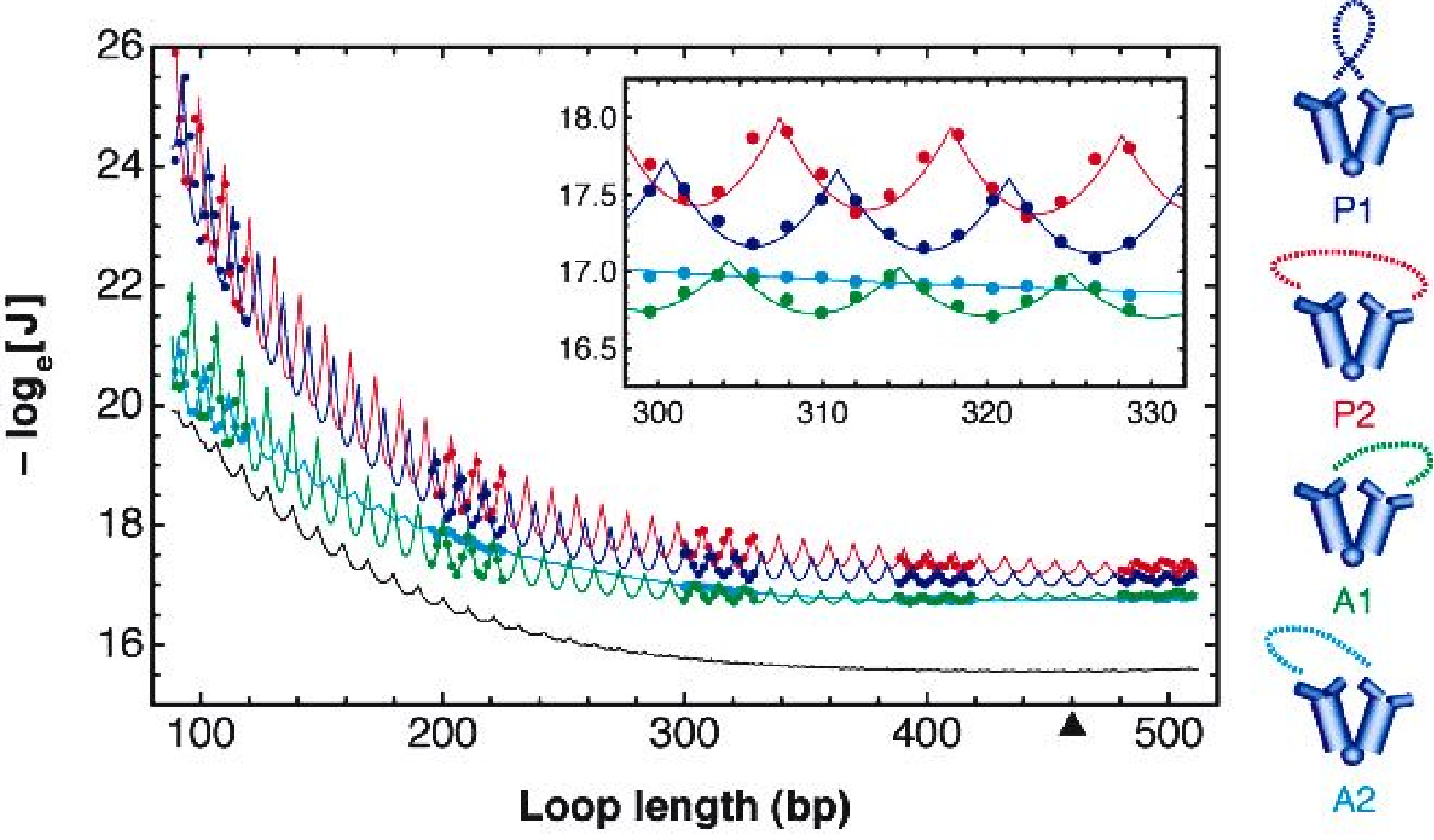}{5}
The $J$ factor for this situation, defined via \erefs{dfJt}--\ref{e:dfJnt}), can be calculated by a simplified version of our Monte Carlo algorithm that generates only the interoperator DNA segments
and hence omits the steric-constraint checking. \Fref{gDNA} shows our calculation of this quantity as a function of loop length. Three sets of Monte Carlo data are reported, each spanning three helical repeats.
The data for each topology are summarized by a global interpolating function equal to the minimum of a collection of parabolas, centered on $\NLl $ values separated by $\ell_{\rm
helix}$. The interpolating functions are specified by the overall phasing (horizontal shift), a scaling function which determines the widths of the parabolas as a function of loop length (physically
representing effective twist stiffness), and an envelope function describing the heights of the successive minima (physically representing competing effects of bend stiffness and entropy). The figure
shows that indeed interpolating functions of this form globally summarize our simulation data over a wide range of $\NLl $ values.  At shorter loop lengths, the contributions of a single
topology seem to dominate at any particular loop length, resulting in a noticeable modulation of the overall looping $J$ factor; however, at longer loop lengths (e.g., $300\,$bp), the contributions of
each topology are all similar and tend to cancel out each others' modulations.  The anti-parallel loop topologies are predicted to be the  preferred state, accounting for
90\% or more of the looped chains for loop lengths of about $89$ to $120\,$bp.

\ifigsize{VilarData}{Comparison of our Monte Carlo results for pure looping to experimental data on \textit{in vivo} repression. Experimental data from {\it in vivo} gene repression experiments
\cite{Muller1996} were converted to $J$ factor values using a formula developed in \Nrref{Saiz2005} (see \aref{SV}) and are shown in \textit{blue}. The \textit{black line} is an interpolation of our
Monte Carlo results and is identical to the one in \fref{gDNA}.}{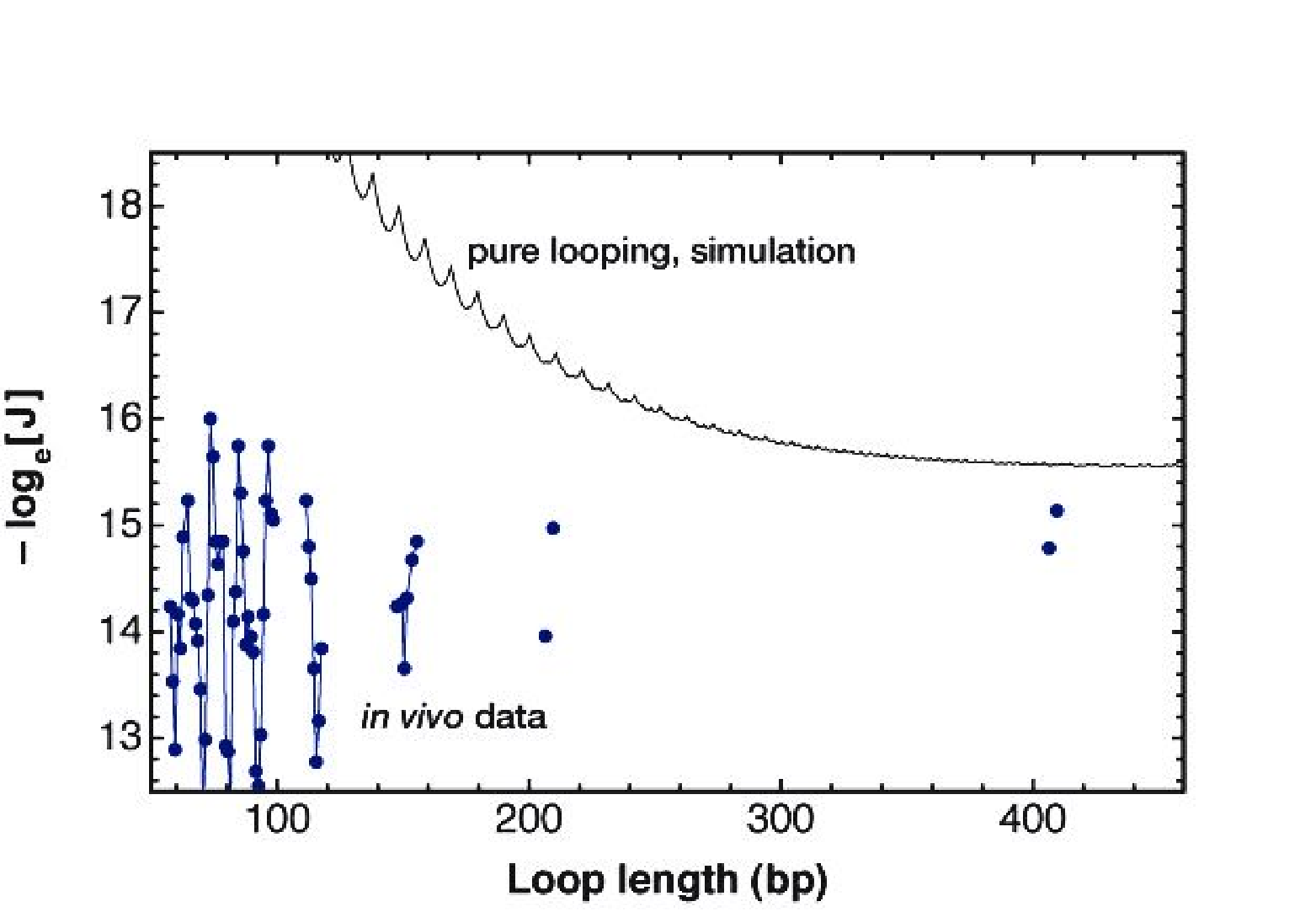}{4.5}
\Fref{VilarData} shows the free energy of looping for an {\it in vivo} repression study \cite{Muller1996}, as interpreted by Saiz et al.\ \cite{Saiz2005}, along with our Monte Carlo results for the
total pure looping $J$ factor. The cellular environment is far from ideal in terms of understanding DNA looping behavior: For example, superhelical stress, other DNA binding proteins, and molecular
crowding all complicate the interpretation. Moreover, some analyses assume that LacI is free in solution at a known concentration \cite{Saiz2005}, whereas much of it is instead likely to be
nonspecifically bound to DNA \cite{bint05b,bint05c} or otherwise unavailable.

Despite these reservations, the comparison to our predictions is interesting: Our calculation seems able to predict the rough magnitude of the \textit{in vivo} looping $J$ factor, to within about a
factor of two, at long loop lengths. At shorter loop lengths, however, \textit{in vivo} looping is far more prevalent than predicted from our simple model. The next subsection presents qualitatively
similar results for the case of \textit{in vitro} TPM experiments.

\subsubsection{TPM looping\pnlabel{s:TPMl}}
For the situation relevant to TPM experiments, the bead and wall must be taken into account. This necessitates use of the algorithm described in \aref{MCa}, to obtain an
estimate of the looping $J$ factor at each loop length and for each topology.  The bead and wall  affected the overall looping $J$ factor, generally reducing it by about 30\% for loops of length
100--300$\,$bp. We can interpret this reduction in terms of the slight entropic stretching force generated by the bead and the wall \cite{sega06a}. We also
found that the presence of the bead and wall significantly changes the relative weights of the various loop topologies from the corresponding pure looping case. For example, consider the ``short''
constructs. Even when the simulation
generates a DNA conformation that qualifies as a type P1 loop, there is some chance that the conformation may be discarded because it violates one of the steric constraints; the chance of retaining a
P1 loop was found to be about twice the corresponding probability for an A1 loop.

We can understand this phenomenon qualitatively as follows: Due to the relatively short length of tether between the wall and the first operator, the DNA is generally
pointing away from the wall when it {\it enters} the loop (at the first operator), thus favoring loops (P1 and P2) that maintain this
directionality and keep the bead away from the wall. This bias is significant because the length of DNA from the second operator to the bead is relatively short. Presumably the reason P1 exhibits a larger shift than P2  is because the P1 topology {\it exits} the loop about $7\,\nmunit$ in front of
where it {\it enters} the loop, whereas P2 {\it exits} about $7\,\nmunit$ behind where it {\it enters}. 

\paragraph{Overall magnitude of $J$}
We first examine the overall magnitude of $J$. To minimize the effects of statistical experimental error, we computed the average quantity $\NJb$ (see \eref{dfJb}) for both the ``long'' and ``short'' constructs. Our Monte Carlo calculation
yielded the value $\NJb(\mathrm{long,\,theory})={100}\, \nMunit$ and the ratio
\begin{equation}\NJb(\mathrm{short,\,theory}) / \NJb(\mathrm{long,\,theory}) = 2.0/{100}\approx{0.020}
.\label{e:shortlongtheory}\end{equation}
That is, harmonic elasticity theory makes the qualitative prediction that the short loop should be strongly penalized for its high elastic energy cost.

Turning next to the experimental values, we faced the problem that TPM data yielding an absolute number for $J$ are so far available only for one loop length (see \aref{data}). However, \eref{relJb}
shows that this one point can be used to normalize all the others. With this procedure, we found that $\NJb(\mathrm{long,\,exp})$ lies in the range $24$--$45\,\nMunit$.
Thus the predicted overall magnitude of the $J$ factor for long loops, computed with no fit parameters, lies  within a factor of {$2.2$--$4$} of experiment, or equivalently our simulation found the free
energy of looping $\NDG$ in agreement with our experimental determination to within about {$\kbt\ln3\approx1\,\kbt$}.

Our uncertainty in overall normalization drops out of ratios such as $\NJb(\mathrm{short,\,exp})/\NJb(\mathrm{long,\,exp})\approx0.35$. Comparing to \eref{shortlongtheory} shows that our theoretical
model cannot account for the relation between short- and long-loop $J$ factors: In this regime, looping is much easier than predicted by our theory. The next paragraph gives more details.

\paragraph{Variation of $J$}
\ifig{JFactorRatio}{Comparison of the relative $J$ factor from our Monte Carlo results (\textit{solid, heavy black curves}) and TPM data of Han et al.\ \cite{han?cLength} on random-sequence DNA
(\textit{open circles with dashed black curves}). \capitem{a}Relative $J$ factors for the ``short'' DNA constructs (see \eref{construct}), based on about $8\ee9$ simulated chains. \capitem{b}Relative
$J$ factors for the ``long'' constructs, based on about $10^{{10}}$ simulated chains. All the experimental $J$ factors are quoted relative to $\NJb(\mathrm{long\,exp})$ defined by \eref{dfJb} for the
experimental data in panel (\textit{b}); similarly, the theory values are relative to $\NJb(\mathrm{long\,theory})$. The blue, red, green, and cyan solid lines represent contributions from P1, P2, A1,
and A2 respectively; the heavy black solid line represents their sum.}{KBT_fig_11.eps}

\Fref{JFactorRatio} shows the behavior of $J$ as we scanned through two ranges of loop lengths (``short'' and ``long''). Because of the large experimental uncertainty in the overall magnitude of $J$, we divided
both theory and experimental values of $J$ by their respective averages $\NJb(\mathrm{long})$, thus forcing both the solid (our theory) and dashed (experiment) black curves in panel~B to  be centered on zero.
\Frefn{JFactorRatio}B shows that, although individual looping topologies have significant phasing effects, these nearly cancel in our simulation results, because in this paper we assume that all four
operator binding orientations have the
same binding energy (see \fref{molecules}). (Similar phenomena were discussed in \Nrrefs{kram88a,saiz07a}.)
\Frefn{JFactorRatio}A shows that the theory embodied by our simulation was unable to account for the relative free energy of looping of long
versus short loops, overestimating $\NDG(94\,\mbox{bp})-\overline{\NDG}(\mathrm{long})$ by up to about ${3.7}\,\kbt$. This observed excess of looping for short DNAs joins other signs of
non-classical elastic behavior, which also begin to appear at short length scales \cite{clou05a,wigg06b}. However, it could instead be explainable in terms of other effects neglected in our model (see
\srefs{ior} and \ref{s:c}).

\subsubsection{Open LacI conformation\pnlabel{s:olrc}}
\Sref{TPMl} showed that the hypotheses of harmonic elasticity,  a rigid V-shaped LacI tetramer, and no nonspecific DNA--repressor interactions,
cannot explain the high looping incidence seen in our experiments for short DNA.
One possible explanation, for which other
support has been growing, is the hypothesis of DNA elastic breakdown
at high curvature \cite{yan04a,wigg05a,wigg06b}. Indeed, \Nrref{wigg06a} showed that such elastic breakdown can accommodate both enhanced looping at short lengths, and normal DNA behavior observed for loops longer than 300$\,\bpunit$.

An alternative hypothesis is that for our shorter loops, looping is actually dominated by the contribution from a distinct, ``open'' conformation of
the repressor tetramer. Accordingly, we repeated our simulation for
one particular representative version of the open conformation, the one discussed in \cite{wong07a}. Here each dimer is assumed to be rigid, but the opening angle of the hinge where the dimers join
has spread to 180$^\circ$. This time we found $\bar J_{\rm loop}(95\,\bpunit)/\bar J_{\rm loop}(305\,\bpunit)\approx 0.13\exp(-\Delta G_{\mathrm{open}})$, where $\Delta G_{\mathrm{open}}$ is the free
energy cost of opening the tetramer. There are a wide variety of estimates of $\Delta G_{\mathrm{open}}$, but we see that even if it were equal to zero, the hypothesis of an open conformation still
would not be consistent with our results.

\section{Effect of looping on bead excursion\pnlabel{s:elbe}}
\Sref{c} will discuss the status of the results in the previous section, but clearly the agreement between theory and experiment is rather rough. We now turn to a much more striking comparison. In
addition to studying the total probability of looping, TPM yields more detailed information about the effect of looping on bead excursion (see \frefs{FitLong}--\ref{f:FitShort}). A common
experimental practice is to bin the data into finite sample windows, giving rise to a probability distribution of bead excursion. In this section we describe how we modeled this
situation theoretically; for more details see \aref{sse}. \Frefs{FitLong}--\ref{f:FitShort} show the degree to which our model successfully predicts the experimental observations.

\subsection{Mimicking looped, doubly-bound DNA tethers\pnlabel{s:mldbdt}}
In the absence of LacI proteins, our procedure is straightforward: We generate chains as in \sref{fsr}, divide them into batches representing $4\,\sunit$ windows (see also \aref{sse}), and compute the RMS
excursion in each batch. Instead of computing the mean of these
$\Nrrmsfour$ values, however, we instead histogram their distribution. We will now apply essentially this same procedure to the
more elaborate calculation of \sref{dl} to obtain the looping distributions we want, with one important modification.

\Sref{dl} considered the {\it potential} for binding the \NOone{}
operator. That is, we computed the fraction of time for which an
unbound \NOone{} operator was positioned close to a pose suitable for
binding to take place. Once binding does occur, however, the geometry
of \NOone{} alters: It develops a kink.
Modeling the bead excursion for looped states requires that we account for the geometry of the fully bound complex, not the about-to-bind state.
({See \aref{ltsls2} for more on this distinction.})

Because we model the LacI tetramer as a rigid object, it may seem that for each selected looping topology we need only consider the DNA outside the loop region, replacing the entire loop by a single
rigid Euclidean motion from the \textit{entry} to the \textit{exit} poses determined by the tetramer (see \fref{CylinderRepresentation}). Eliminating the loop region from the simulation would
certainly speed up calculations, and indeed, it is nearly correct. However, this procedure would miss the possibility of steric clashes between the loop region and the bead and wall, potentially skewing the
reported distribution of bead excursions. As a compromise between speed and accuracy, we simulated the regions {wall}$\to$\textit{entry}, and {\textit{exit}}$\to${bead}, as
usual, but, for each looping topology, inserted a {\it representative} loop between them. The representative was an actual loop configuration stored from the more exhaustive simulation in \sref{dl}. The entire chain was then
checked for steric clashes as usual, and the distribution of bead excursions for each of the four looping topologies was built up.

To find the appropriate relative weights to assign each of these four distributions, we used the simulations described in \sref{dl}. Finally, we combined the resulting overall distribution for looped
states with the corresponding one for the unlooped state. In principle, we could have found the appropriate relative weighting by using our computed looping $J$ factor and the binding constant $\NKd$
extracted from experimental data in \sref{edlJf}. In practice, however, the experimental data do not yet yield very precise values for $\NKd$. Moreover
the theoretical prediction for overall weighting depends very sensitively on the value of DNA persistence length, which is not precisely known.  Accordingly (and in
the spirit of \fref{JFactorRatio}), we chose the overall relative weighting between looped and unlooped states by hand for one value of $\NLl$. That is, we chose
a common, constant value of this factor for all curves shown in \fref{FitLong}.  Our justification for this step is  that
our adjustment does not modify the locations, widths, nor \textit{relative} strengths of the looped-state peaks, which are thus no-parameter
predictions of the theory.\Nfootnote{}{More
precisely, we started with the predicted bead excursion histograms $P_{\rm loop}(\rho)$ and $P_{\rm noloop}(\rho)$. Then we chose a constant
$\lambda$ and displayed $(\lambda P_{\rm loop}+P_{\rm
noloop})/(1+\lambda)$. Choosing a value for $\lambda$ that is smaller
than unity thus enhances the relative contribution of the unlooped states.}

\subsection{Bead excursion results\pnlabel{s:RMSr}}
\ifigsize{FitLong}{Theory and experiment for the probability density functions of RMS bead excursion for six of the ``long chain'' constructs ($L\approx900\,$bp, $\NLl\approx306\,$bp) studied by Han
et al.\ \cite{han?cLength}.
\textit{Blue dashed curves} show experimental TPM data on random-sequence DNA. \textit{Black curves} show our theoretically predicted distributions for the corresponding interoperator spacings.
The model yields these histograms as the sum of five contributions, corresponding to the four looped topologies  (A1, A2, P1, and P2), and the unlooped state.
Because our simulation results were not fits to the data, they did not reproduce perfectly the ratio of looped to unlooped occupancies. For visualization, therefore, we have adjusted this overall ratio by a factor common
to all six curves (see main text). This rescaling does not affect the locations of the peaks, the relative weights of the two looped-state peaks, nor the dependences of weights on loop length $\NLl$, all of which are
zero-fit-parameter predictions of our model. The separate
RMS displacements for each individual loop topology, for the $300\,$bp case, are also shown as \textit{vertical dashed lines}.}{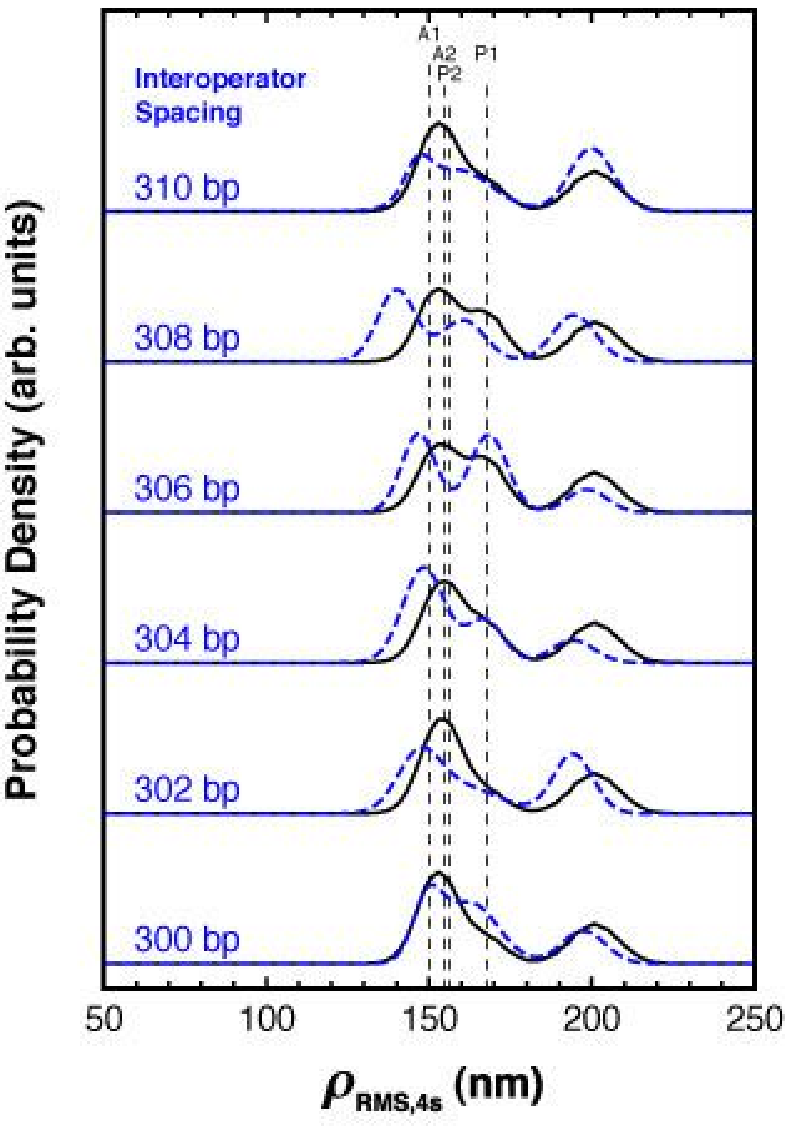}{2.9}
\ifigsize{FitShort}{Theory and experiment for the probability density functions of the finite-sample RMS bead excursion for our three ``short chain'' constructs. The separate
RMS displacements for each individual loop topology, for the $89\,$bp case, are also shown as \textit{vertical dashed lines}. In other respects the figure is similar to \fref{FitLong}.
 }{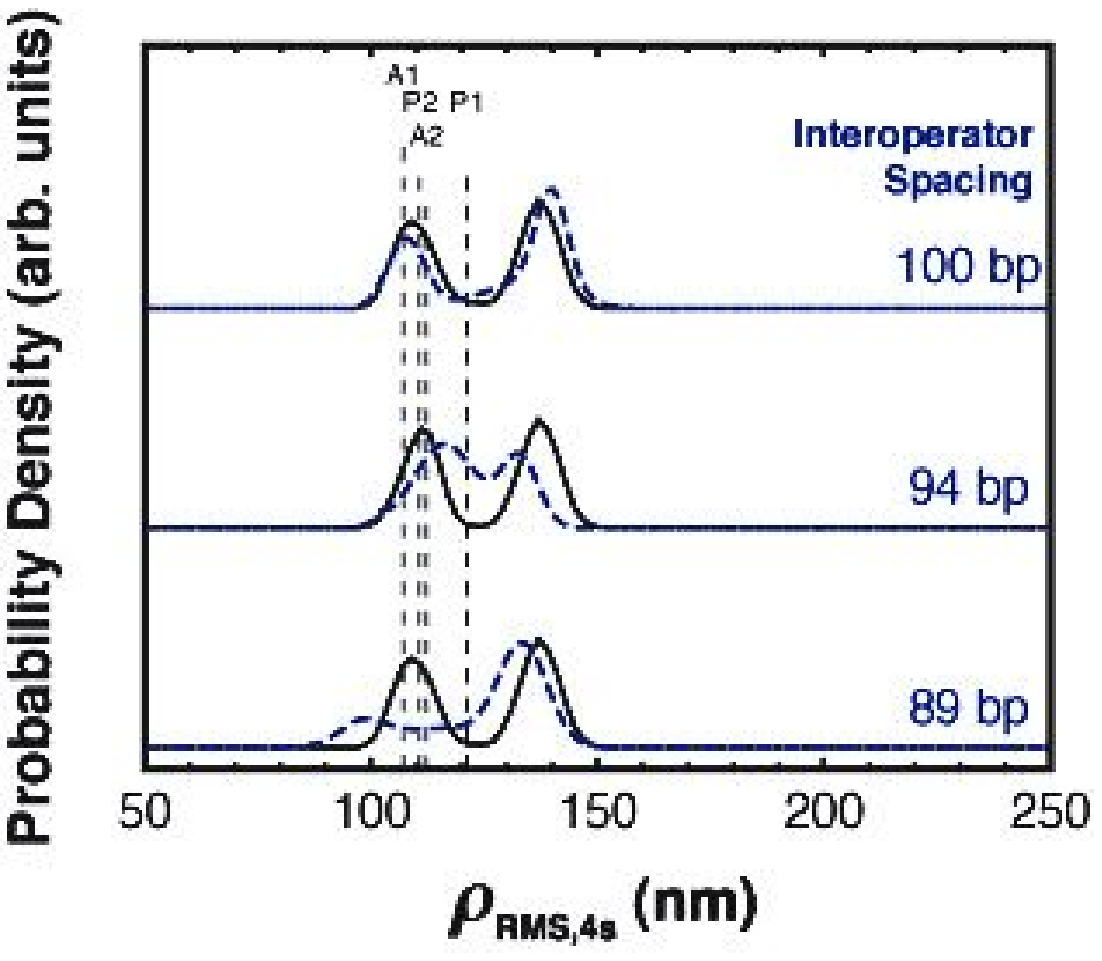}{2.9}

We followed the method of the previous subsection, including applying a modified blur correction appropriate for looped tethers (see
\aref{etlbc}).
In order to obtain smooth distributions, we ran the Monte Carlo code until a total of $2.5\times 10^4$ observations (independent values of $\Nrrmsfour$) were obtained.  The results were then binned with bins of width $2\,\nmunit$
and normalized.

\Frefs{FitLong} and \ref{f:FitShort} show the experimentally determined probability density functions for bead excursion as blue dashed lines. The rightmost peaks in these distributions correspond to our
expectations for unlooped tethers (\fref{calibrplot}). One might think that at least one of the remaining peaks would be located at a value corresponding to an imagined tether that is unlooped, but
shortened by the number of basepairs between the two operators; on the contrary, this expected peak location generally falls \textit{between} the two peaks seen in the data \cite{han?cLength}.
\Frefs{FitLong}--\ref{f:FitShort} show that in contrast, our simulation does a fairly good job of predicting \textit{both} peak locations.
Indeed, the figures show significant resemblance between the theory and experiment, including the trends as $\NLl$ is varied. Specifically, the theory automatically yields two
looped peaks, and roughly gives their observed locations and widths (each to within about $10\,\nmunit$). The theory also predicts that the middle peak is small at $302$--$304\,$bp, and bigger elsewhere, and
that the lower peak is not modulated as much as the middle one, all of which are in agreement with the experimental data. All of these qualitatively satisfactory conclusions emerge without the hypothesis of
any major conformational change of LacI.

The dissection of bead excursion into distinct peaks is also relevant to the question of a possible open conformation of the LacI tetramer. Even if we suppose that the middle looped peak in
\fref{FitShort} reflects an open LacI conformation, as proposed by Wong et al., nevertheless those authors also proposed that the lower peak reflected the closed (V-shaped) conformation \cite{wong07a}. The
experiments of Han et al.\ show that these peaks have comparable strength, and so in particular the lower one is inconsistent with the assumption of harmonic DNA elasticity. (One could
instead propose that \textit{both} peaks reflect an open conformation, but \sref{olrc} argued that even this new hypothesis is unlikely to explain the experimental results.)

\section{Relation to others' results\pnlabel{s:rtor}}
The calculational approach in this paper is Gaussian sampling Monte Carlo  (see \sref{ctdwl}).
Here we make just a few specific comments about representative examples of other calculational methods. The reader may wish to pass directly to the discussion in \sref{c}.

\subsection{Analytic methods}Some methods do not involve the generation of random matrices:
\paragraph{Diffusion equation on $E(3)$:}A series of independent steps defines a random walk.  Thus the successive bends between chain segments can be regarded as defining a walk on the
group manifold of $SO(3)$, the rotation group; more generally, the successive poses of segments define a walk on the Euclidean group $E(3)$. The probability density function of segment poses
evolves as we move along the chain, in a way that can be calculated by using a set of orthonormal functions \cite{chir00a,yan05a}, a procedure mathematically similar to some calculations in quantum
mechanics.
In some cases the resulting series can be summed and represented as a continued fraction \cite{spak04a,spak05a,spak06a,mehr08a}. Another approach to the evolution of a distribution is via matrix
exponentiation \cite{doua05a}, or other transfer-matrix approaches \cite{ranj05a}.  These approaches converge  slowly, however, for the case of short chains, and none accommodates easily the sort of nonlocal constraints arising in TPM experiments.
\paragraph{Gaussian approximation:}The looping $J$ factor is essentially a probability for configurations satisfying a global constraint. As such it can be represented as a functional integral, which
in turn can be approximated by an expansion of its integrand around its critical points  (maxima). Keeping the value of the integrand only at the critical points recovers the equations
of elastic rod equilibrium; however, it omits entropic contributions to the free energy (see for example \cite{Geanacopoulos2001,puro06a,goya07a}). An improvement to this procedure approximates the integrand as a
sum of  Gaussians about its maxima; the integral may then be
done, yielding the entropic contribution as the log of a functional determinant \cite{shim84a,zhan03a,kuli05a,zhan06a,zhan06b,swig06b}. Unfortunately, this approximation breaks down when any eigenvalue of the fluctuation
operator becomes small, and in particular when the loop becomes bigger than a few hundred basepairs. Also, it is again difficult to generalize this approach to incorporate nonlocal constraints such as bead--wall
exclusion. And although the method is efficient for comparing the free energies of different looped topologies, the direct comparison of looped to unlooped is difficult. Perhaps for this reason, we do not know of any work using this approach that has attempted to include the
dependence on LacI concentration; that analysis was crucial to the present work in order to disentangle the effects of $J$ and $\NKd$ in the experimental data.

Note, however, that Zhang et al., using this approach, have introduced a more detailed model of the LacI conformation than the one in the present paper \cite{zhan06a,zhan06b}.
Also, they and Swigon et al.\ introduced a detailed model of sequence dependence in their calculations, unlike the present work \cite{swig06b}.\Nfootnote{fnj}{Popov and Tkachenko also studied statistical
properties of sequence-dependence effects \cite{popov05b} in models of this type. See also \Nrrefs{mann96a,olso04a,rapa04a,goya07a,zico08a}.} Like the present work, \Nrrefs{zhan06a,zhan06b,swig06b} neglected
possible ``wrapping'' effects. Swigon et al.\ considered very low salt
concentrations (and so had to account for long-range electrostatic effects), so their results cannot be directly compared to ours; moreover they considered only the situation we have called ``pure
looping'' (\sref{Jfpl}). Nevertheless, it is interesting that for pure looping,
our results agree qualitatively with their finding that, assuming the V-conformation of LacI, the antiparallel loop has lowest looping free energy.

\subsection{Monte Carlo methods}We chose a Monte Carlo method because it is computationally tractable, it calculates the quantities actually observed in TPM experiments, it
covers the entire range of loop lengths of interest to us, and it allows a simple implementation of all steric
constraints relevant to our problem. Additionally, the method can readily be generalized to include sequence dependence \cite{czap06a} or nonlinear models of DNA elasticity \cite{wigg06a}.
As a bonus, Monte Carlo methods give a direct visualization of which
chain conformations, and in particular which topologies, dominate the statistical sum, unlike the diffusion equation or matrix-exponentiation methods.
It also gives us the distribution of near-miss configurations, allowing us to quantify the importance of the stereospecificity of binding (\sref{odls}). Also, we
are not obliged to find every critical point of the elastic energy, an error-prone search in a high-dimensional space of configurations. For example, it is easy to miss stability bifurcations as the
loop length is stepped through a range. Our Monte Carlo code stumbles upon the dominant configurations, including all topologies, without requiring human insight. (It also automatically lumps together all
distinct topologies that cannot be experimentally distinguished, without the need to find their individual critical points, then add the corresponding contributions by hand.) Finally, unlike some
methods, Monte Carlo simulation easily allowed us to work out the distribution of bead excursions (\sref{elbe}).

Among prior Monte Carlo methods we mention:
\paragraph{Gaussian Sampling:}The work in this paper extends prior work in \Nrrefs{sega06a,nels06a}.
As mentioned before, Czapla et al.\ also applied this method to cyclization (but not looping) \cite{czap06a}. \Aref{emcz} makes a specific point about a side calculation in that work.
\paragraph{Metropolis Monte Carlo, Brownian Dynamics:}These powerful and general methods can in principle handle chains of any length, with arbitrarily complex constraints, and in some forms can also
study kinetics (see, e.g., \Nrref{hage85a,leve86a,merl98a,volo96a,podt99a,podt00a,zhan00a,huan01a}). We only note that for the equilibrium calculations of interest to us, Gaussian sampling Monte Carlo
is a simpler alternative method, which completes in a reasonable
time on current laptop computers. Moreover, because in GS Monte Carlo every chain is independent of every other one, we have fewer worries about pre-equilibration, coverage of the allowed phase space, and so on than in
Markov-chain MC methods.
\paragraph{All-atom simulation:}Schulten and couthors have developed a hybrid method that represents the DNA as an elastic continuum, coupled to an all-atom simulation of the LacI tetramer
\cite{bala99a,vill04a,bala04a,Balaeff2006,Villa2005}. They did not apply this method to TPM experiments, and their simulation appears to neglect twist--bend coupling and entropic contributions from
the DNA chain. However,  simulations of this type did identify a ``locking'' mechanism, which apparently prevents opening of the tetramer (i.e.\ maintains its overall V-form), even in the presence of significant external stress.

\subsection{Fitting}
Some insight into mechanisms can be gleaned by fitting data to phenomenological parameters roughly representing DNA stiffnesses etc., essentially obtaining interpolation formulas summarizing
the data \cite{Saiz2005,saiz07a}. For example, the anomalously high looping compared to theoretical expectations (\sref{Jfpl}) was previously noted in \Nrref{Saiz2005}, and the possibility of
partial cancellation of phasing modulation (\sref{Jfpl}) in \Nrref{kram88a,saiz07a}.
\NRref{Saiz2005} also called attention to an asymmetric modulation in their graphs of looping free energy versus $\NLl$; we agree with their later observation that such asymmetries can arise from
the sum of different loop topologies  (fine structure in \fref{VilarData}).

A drawback of the fitting approach, however, is that the inferred values of
fit parameters do not have a literal interpretation as elastic constants; for example, their numerical values depend on extraneous variables \cite{saiz07a}.
The present work sought instead to predict the data from first principles, using fixed values of elastic constants. Also, many prior works studied \textit{in vivo} data, whereas we have focused on TPM
experiments for reasons described earlier.

\subsection{Estimates and other analyses}The analysis of Wong et al.\ attempted to estimate the positions of the peaks in their TPM data from simple geometry applied to assumed configurations for the repressor tetramer
\cite{wong07a}. The present work sharpens and corrects some qualitative estimates made, for example, in their Supplement.

Vanzi et al.\ obtained a looping $J$ factor $\approx 0.1\,\nMunit$ from their data and noted that this value is much lower than those measured in other types of experiments \cite{vanz06a}. Both our
theory and our experimental results obtained from data in \Nrref{han?cLength} agree that $J$ is much larger than $0.1\,\nMunit$, albeit with large uncertainties for now. Vanzi et al.\
suggested that it would be important to calculate effects such as the entropic force generated by the bead;  the present work gives the needed calculations.

\section{Discussion\pnlabel{s:c}}
Our theoretical model and its main results were summarized in \sref{rtp}.

As mentioned earlier, DNA cyclization and looping have been the focus of many prior calculations. Broadly speaking, the novelty of the present work lies in the combination of a number of features: We
have calculated looping, not cyclization; we have calculated quantities relevant to TPM experiments; and we have presented detailed comparisons to experimental data, with
no fitting parameters. (\Sref{stel} explained why we found TPM to be a particularly revealing class of experiments.) Finally, we know of no prior work that predicts the detailed distribution of bead
excursions in looping as functions of loop length and LacI concentration. TPM experiments yield such data, and with it the prospect of experimentally separating the contributions of different loop
topologies.

\Fref{calibrplot} shows that our simple model adequately captures much of the physics of equilibrium tethered particle motion without looping. The remaining discrepancy between theory and
experiment may reflect unremoved instrumental drift. Alternatively, the effective bead size may be slightly different from the nominal value, or effectively different due to surface irregularity. 
Despite these reservations, clearly our understanding of TPM is more than adequate
to distinguish changes in effective tether length of $100\,\bpunit$ or more.

\Sref{Jfr} gave a determination of the absolute magnitude of the looping $J$ factor as a function of operator spacing, and a comparison to experiment. These comparisons were only approximately
successful. We may point out, however, that the experimental determination had large uncertainties, because the available data are still somewhat sparse. In particular, all our absolute values
currently depend on a single measurement, of the fraction of time spent looped for a $306\,$bp loop at $\Nlaci=100\,\pMunit$ (see \eref{relJb}). Also the slope of the titration curve, and hence the
parameter $J_*$, is still poorly characterized by available data (see \fref{titrate} and \eref{relJb}). Additional experiments would help improve this situation.

On the theory side, we note that existing measurements of the DNA persistence length $\xi$ are subject to uncertainties, and that rather small changes in the assumed value of $\xi$ make significant
changes in our prediction for the overall $J$ factor. Thus an accurate absolute prediction of $J$ is perhaps demanding too much at this time; and in any case we also just argued that
experiments, too, do not yet yield an accurate absolute experimental determination of $J$. To address both of these concerns, we noted that the \textit{relative} $J$ factors for different situations are
better determined by experiments than the absolute magnitude, and so we  scaled the theory and experimental results by their respective averages for loop lengths near $305\,$bp. Then we compared the predicted 
and observed relative $J$ factors
for several individual lengths near $305\,$bp, and also for a few lengths near $94\,$bp. Here the results were that \textit{(i)}~near  $305\,$bp,
neither theory nor experiment were strongly modulated by phasing, though experiment was more modulated than theory (\frefn{JFactorRatio}B); \textit{(ii)}~near $94\,$bp, our theory predicts far less
looping (lower $J$) than was observed (\frefn{JFactorRatio}A).

The fact that our model underestimates $P_{\rm loop}$ for short loops cannot simply be attributed to the model's neglect of sequence information. Indeed, special sequences are observed to cyclize
\cite{Cloutier2004} and to loop \cite{Mehta1999,morg05a,han?bSeque} even more avidly than the random-sequence constructs studied here and in \Nrref{han?cLength}. Nor can we simply suppose that we
overestimated the true value of the DNA persistence length; reducing the value in the simulation would increase looping at both short and long lengths, leading to a worse discrepancy at the long end.
Instead, we must look to our other physical hypotheses to see which is breaking down for short DNA (see \sref{ior}).

The very small phasing modulation observed in our calculations results partly from the near cancellation of out-of-phase modulations in the contributions from individual looping topologies. Certainly
we could have obtained greater modulation had we been willing to adjust our model's twist stiffness \textit{ad hoc,} but our goal was to see how well the model predicted the data without fitting. It is
possible that the degeneracy we assumed between the binding in each of the four topologies is an oversimplification, and that therefore the cancellation is not as complete as what we found in our
simulation. (For example, we neglected the strain on the tetramer exerted by the DNA, which could differentially affect the different looping topologies.)

Our results were much more striking when we turned to the detailed distributions of bead excursions. These results are complementary to the ones on absolute and relative $J$ factors, because the
locations of the peaks are not affected by any possible elastic breakdown of the DNA within the loops;  instead, they depend sensitively on the assumed geometry of the repressor tetramer.
\Fref{FitLong} shows that the distinctive three-peak structure, including the
positions, and even the relative strengths, of the two lower peaks, emerges as a natural consequence of the four discrete looping topologies for LacI and its known geometry.
We also found reasonable agreement with the less extensive available data on the peak positions for the short-loop bead excursion distribution (see \fref{FitShort}).

Our calculations did not systematically study alternate, ``open,'' conformation of LacI such as the one proposed in \Nrref{edel03a}, although we did simulate one somewhat artificial model (a rigid,
$180^{\circ}$ conformation \cite{wong07a}). Although certainly such a conformational switch is possible, we note that
Villa et al.\ have argued against it on grounds of molecular mechanics \cite{Villa2005}. On the other hand, Wong et al.\ observed \textit{direct} transitions between their two looped states (i.e.,
without an intervening unlooped episode), and they argued that this rules out any interpretation of the different looped states in terms of the distinct binding topologies. Our work has not resolved
this issue. But in any case,  the four loop topologies we studied must exist with any LacI conformation, and we showed that the closed conformation alone gives a surprisingly detailed account of the
observations of Han et al.\ \cite{han?cLength}. We also found that when we simulated the maximally advantageous, $180^{\circ}$, conformation, the resulting looping $J$ factor still fell short of
the value observed in experiments, even if we supposed zero free energy cost to pop into that state.
And if desired, our calculation scheme may easily be extended to incorporate any desired hypothesis for LacI opening.

Finally, to illustrate the generality of our method, we also calculated $J$ factors for a DNA with no bead or wall (``pure looping''). This calculation also gave us a quantitative estimate of the
effect of the bead and wall on looping.

\subsection{Future experiments\pnlabel{s:fe}}
More extensive titration experiments will help pinpoint the values of $J$ better, and eliminate the need to base all determinations of $J$ on a single titration curve, as we have been
compelled to do. Also, taking data with a fast camera shutter would eliminate, or at least minimize, the role of the blur correction (see also \aref{bc2}).

Our ability to resolve bead excursion distributions into distinct contributions from different looping topologies in \sref{elbe} involved a subtle tradeoff.
The finite sample RMS bead excursion, $\Nrrmst$, is more sharply peaked for longer sampling time $t$. Thus, using larger values of $t$ could in principle clarify the structure of the distributions
in \frefs{FitLong}--\ref{f:FitShort}. But increasing $t$ also increases the fraction of incidents when a tether changes its looping state in the middle of a sample. One could imagine instead increasing
the video frame rate, but \aref{sse} points out that the bead diffusion time sets a limit to what can be gained in this way. Thus it would be desirable to do experiments with smaller beads, hence
faster bead diffusion times, and correspondingly faster video frame rate.

A second advantage to using smaller beads is that this would minimize
the perturbation to looping caused by the bead, for instance the
entropic force pushing the bead away from the wall, which slightly
stretches the DNA.

Finally, it would be interesting to try experiments of the sort considered here but using other regulatory proteins, particularly ones not suspected to be as labile as LacI, in order to see if the effects
we calculated really are generic, as we believe they are.

\subsection{Future theory\pnlabel{s:ft}}
Certainly we could improve agreement with experiment by introducing two fitting parameters, which could be considered as effective twist and bend stiffnesses.\Nfootnote{fnl}{The other two elastic
constants, involving anisotropy and the twist--bend
coupling, had small effects on our results (see \aref{ciem}).} Alternatively, the method in this paper could trivially be adapted to a detailed model of sequence-specific, harmonic DNA elasticity.
But such detailed models may miss a larger point, that harmonic elasticity itself seems to break down at high bend and/or twist strain. An advantage of our Monte Carlo scheme is that it does not
depend on the assumption of a Gaussian distribution of elementary bends; any distribution of interest may be substituted in the code, for example the one proposed in \Nrref{wigg06b}. Finally, any
desired characterization of repressor flexibility, for example the one outlined in \Nrref{wong07a}, can be incorporated by drawing the matrices $\NM,\ \NN$, etc.\ (see \aref{ltsls2}) from appropriate
distributions.

\subsection{Conclusion\pnlabel{s:concl}}
Our goal was to go the entire distance from an elasticity theory of DNA to new, quantitative experimental results. To cast the sharpest light on the comparison, we chose to study experiments that are free
from confounding elements present \textit{in vivo}, e.g., molecular crowding and DNA bending proteins other than the repressor of interest. We developed a number of techniques that will also be relevant for other experiments
involving tethering of particles near surfaces. Although even this simplified setting presents some daunting geometrical complications (the effects of the tethered bead and wall), nevertheless it
really is possible to understand much of the available experimental
data (e.g., the three-peak structure of the bead excursion
distribution) with a physically simple model. This level of detailed
agreement will be helpful when trying to identify the many peaks in future
experiments involving more than two DNA-binding sites. It also gives
us strong evidence that our experiments are working as intended; for
example, if bead sticking events were corrupting our data, it would
be quite a coincidence if nevertheless we found agreement with theory
for the full histogram of bead excursions. We did find
significant deviations between theory and experiment, at short loop
lengths. Here the fact that our underlying physical model had very few assumptions lets us focus attention more specifically on what modifications to those
assumptions may be needed.

\ack
{We thank Paul Wiggins for suggesting the general scheme of the calculations, and
Kate Craig,
Nily Dan,
Jeff Gelles,
Lin Han,
Stephanie Johnson,
Mitch Lewis,
Ponzy Lu,
Davide Normanno,
Gasper Tkacik,
and
Elizabeth Villa
}for many helpful discussions.
PN, JB, and KT were supported in part by NSF grants DGE-0221664, DMR04-25780, and DMR-0404674.
RP acknowledges the support of the Keck Foundation, National
Science Foundation grants CMS-0301657 and CMS-0404031, and the
National Institutes of Health Director's Pioneer Award grant  DP1
OD000217. HG was supported in part by both the NSF funded NIRT
and the NIH Director's Pioneer Award.

\section*{References}
\bibliographystyle{unsrt}
\bibliography{montecarlo}

\end{document}
\end{document}